\documentclass[prb,aps,twocolumn]{revtex4}

\usepackage{times}
\usepackage{graphicx}
\usepackage{float}
\usepackage{latexsym,amsmath,amssymb,bm,euscript}
\usepackage{color}
\usepackage{subfigure}
\usepackage{epstopdf}
\usepackage[colorlinks=true,linkcolor=blue,citecolor=blue]{hyperref}
\usepackage{type1cm}
\usepackage{ulem}
\usepackage{cancel}
\usepackage{tikz}
\usetikzlibrary {positioning}
\usepackage{appendix}

\begin{document}
\title{Abelian origin of $\nu=2/3$ and $2+2/3$ fractional quantum Hall effect}

\author{Liangdong Hu$^{1,2,3,4}$, W. Zhu$^{2,3,4}$}

\affiliation{$^1$Zhejiang University, Hangzhou 310027, China}
\affiliation{$^2$School of Science, Westlake University, 18 Shilongshan Road, Hangzhou 310024, Zhejiang Province, China}
\affiliation{$^3$Institute of Natural Sciences, Westlake Institute for Advanced Study, 18 Shilongshan Road, Hangzhou 310024, Zhejiang Province, China.}
\affiliation{$^4$Key Laboratory for Quantum Materials of Zhejiang Province, School of Science, Westlake University, 18 Shilongshan Road, Hangzhou 310024, Zhejiang Province, China.}

\begin{abstract}
We investigate the ground state properties of fractional quantum Hall effect at the filling factor $\nu=2/3$ and $2+2/3$, with a special focus on their typical edge physics.
Via topological characterization scheme in the framework of density matrix renormalization group, the nature of $\nu=2/3$ and $2+2/3$ state are identified as Abelian hole-type Laughlin state, as evidenced by the fingerprint of entanglement spectra, central charge and topological spins. Crucially,	
by constructing interface between $2/3$ ($2+2/3$) state and different integer quantum Hall states, we study the structures of the interfaces from many aspects, including charge density and dipole moment. In particular, we demonstrate the edge reconstruction by visualizing edge channels comprised of two groups: the outermost $1/3$ channel and inner composite channel made of a charged mode and neutral modes.
\end{abstract}

\date{\today}

\maketitle
\tableofcontents

\clearpage
\section{Introduction}

The fractional quantum Hall (FQH) \cite{Tsui1982,Laughlin1983} effect is the foremost example of topologically ordered state \cite{Wen1990}, which attracted a lot of attention since its discovery. The FQH state exhibits novel features of topological order, including the fractional charge excitations \cite{Laughlin1983}, anyonic statistics and the topologically protected edge property \cite{Wen1992}. Especially, the study of edge physics is in the central stage of the FQH effect. First, due to the well-established bulk-edge correspondence, the knowledge of edge physics, which is more accessible in transport experiments, could peek into the topology in the bulk. In this context, robustness of transport properties such as quantization of electrical and thermal conductance, can all be intuitively understood by the formation of edge states. Second, beyond the description of topological order in the bulk, interplay between strong interactions and confining potentials may ``reshape" the FQH edge states, leading to phenomenon of the edge reconstruction  \cite{Kane1994,Chamon1994,Wan2002,MacDonald1993}, which introduces complications into the edge physics.
This also motivates recent experimental progresses to detect the emergent neutral modes \cite{Bid2009,NP2/3_2017,Cohen2019,Grivnin2014,NovelDutta2021}.
To sum up, the edge physics, in the presence of tunneling and disorder, is an enduring theme in the study of FQH effect.

In this paper, we will consider the FQH state at the filling factor $\nu= 2/3$ in the lowest Landau level.
The spin polarized  $\nu= 2/3$ FQH state is usually regarded as a particle-hole-conjugated Laughlin $\nu=1/3$ state or, equivalently,
a hole $1/3$ state embedded in the  integer quantum Hall $\nu=1$ background \cite{Meir1994}. This is the simplest example with counter-propagating edge modes, thus the nature of $2/3$ state has been widely studied both theoretically and experimentally \cite{Meir1994,Kane1997,Tapash1999,Shibata2007,Peterson2015,Davenport2012,WPan2012,Eisentein1990,Shayegan1992,Smet2001,Levy2001,Haug2004,Sorba2007}.
Especially, the upstream neutral edge mode has recently been identified in experiments \cite{Bid2009,NP2/3_2017,Cohen2019,Grivnin2014,NovelDutta2021}, together with a  more delicate indication of edge reconstruction \cite{Zixiang2008,Yinghai2012,Gefen2013,Gefen2014}.
One of our motivation is to provide clear evidence of possible edge reconstruction of $2/3$ state and uncover its connection with the experimental observations \cite{Bid2009,NP2/3_2017}.

Additionally, we will also consider the FQH state at the filling factor $\nu=2/3$ in the second Landau level (refer it as $2+2/3$ or $8/3$ state).
Owing to the possibility of non-Abelian states in the second Landau level,
the $2+2/3$ state may potentially realize non-Abelian type topological order, such as
the $Z_4$ parafermion FQH state \cite{Read1999,Barkeshli2010}, the Fibonacci $SU(3)_2$ state \cite{Vaezi2014}, and the Bonderson-Slingerland state \cite{Bonderson2008}.
To our best knowledge, the precise nature of edge physics of $2+2/3$ state and the trace of non-Abelian predictions have not been studied systematically before  \cite{Peterson2015,Davenport2012}, which is another motivation of this work.

In this paper, we present a microscopic  study of the $\nu=2/3$ and $2+2/3$ state
 based on the cylinder geometry, from the viewpoints of bulk and edge properties.
First of all, through a topological characterization scheme for bulk properties, we identify that both $2/3$ and $2+2/3$ states are topologically equivalent to a hole-type Laughlin state with Abelian statistics, based on the numerical evidences from the entanglement spectra and momentum polarization. 
Apart from the topological characters in the bulk, we are able to study the topological order by investigating the interface between $2/3$ ($2+2/3$) and different integer quantum Hall (IQH) states, and these man-made interfaces provide a method to study a variety of edge as well as interface physics.
In particular, we find strong signals for edge reconstruction on $\nu=0|\nu=2/3$ and $\nu=2|\nu=2+2/3$ interface, and we identify  the outer  and inner edges are located at density changes of $0\rightarrow 1/3$ and $1/3\rightarrow 2/3$, respectively.
These results not only test predictions of the effective-field theory,
but also gain further insight into the physics of the edge reconstruction.

Moreover, we contribute two different methods to obtain the topological shift quantum number \cite{Zee1992}, which provides an ambiguous information for $ \nu=2/3,2+2/3$ state.
One method is through the momentum polarization from entanglement spectra \cite{Zaletel2013,MomentPolar}, which is based on a uniform bulk state.
The other method is via electric dipole momenta living on the FQH edge or interface \cite{YJPark2014}.
In particular, both of these two methods give the same topological shift, where $\mathcal{S}=0$ for $\nu=2/3$ state
and $\mathcal{S}=2$ for $\nu=8/3$ state. In Ref.\cite{note1}, the theoretical topological shift of a fully polarized Abelian $\nu=2/3$ state
in higher Landau level(i.e $\nu=n+2/3 (n=0,1,2,\cdots)$) is $\mathcal{S}=2n$. This provides another evidence of Abelian nature for these two states.

The rest of the paper is organized as follows. In Sec.\ref{sec::Model}, we will illustrate the details of implementing DMRG in FQH system and the construction of interface in DMRG. We will summarize the quantum numbers that determine the topological order in the bulk, on the edge of FQH states. As results, in Sec.\ref{sec::bulktopo}, we will show the topological charaters of $\nu=1/3,2/3,8/3$ states. Finally, in Sec.\ref{sec::edge} and Sec.\ref{sec::interface}, we will study the edge, and interface physics between FQH($\nu=1/3,2/3,8/3$) and IQH states($\nu=0,1,2,3$).

\section{Model and Method}\label{sec::Model}
\subsection{Model}
Throughout this paper, we consider interacting electrons living on the cylinder geometry in the presence of a uniform perpendicular magnetic field. The cylinder has area $S=L_x L_y$,
where $y$ runs along the periodic direction with circumference $L_y=L$ and $x$ runs along the open direction with length $L_x$.
In the Landau gauge $\bm{A}=(0,-Bx)$ and magnetic field $\bm B = -\bm{e}_z B$ (or $B_z=-B$, $\bm{e}$ is the unit vector), the single particle orbital in the $n$th-Landau level($n$th-LL) is
  \begin{equation}\label{LL-cylinder}
 \psi_{n,j}(\bm r) = \frac{\mathrm{e}^{i \frac{X_j}{\ell^2} y-\frac{(x-X_j)^2}{2\ell^2}}}{\sqrt{2^n n! \sqrt{\pi} \ell L_y}}
 H_n\left(\frac{x-X_j}{\ell}\right),
 \end{equation}
where $H_n(x)$ is the Hermite polynomial, $\ell=\sqrt{\hbar/|eB|}$ is magnetic length and $X_j=k_j\ell^2$ is the
center of single particle orbit. There are $N_\phi$-fold degenerated orbitals in a single Landau level $2\pi\ell^2N_\phi=L_x L_y $, which are distinguished
by the momentum quantum number $k_j=\frac{2\pi j}{L_y}(j=0,1...,N_\phi-1)$.
When a single LL is partially occupied with a  fractional filling factor $\nu=N_e/N_\phi=p/q$ ($p$ and $q$ are integers and coprime to each other),
one can consider the electron-electron interaction by projecting onto the $N$th Landau levle(LL):
 \begin{equation}\label{oriH}
 \hat H_I = \sum_{j_1,j_2,j_3,j_4}^{N_\phi} A^N_{j_1,j_2,j_3,j_4} \hat c^\dagger_{j_1}\hat c^\dagger_{j_2}\hat c_{j_3}\hat c_{j_4}.
 \end{equation}

 In this paper, we adopt two types of interactions, Haldane pseudopotential and modified Coulomb interaction $V(r)=\frac{e^2}{\epsilon r}\mathrm{e}^{-r^2/\xi^2}$ with a regulated
 length $\xi=4\ell$~\cite{Zaletel2015}. More details about the derivation of $A^N_{j_1,j_2,j_3,j_4}$ can be found in Appendix \ref{apdx:Hamiltonian}, we only show the result here
 \begin{align}\label{coef}
 \nonumber
 A^N_{j_1,j_2,j_3,j_4} &=   \frac{1}{2L_y}\int_{-\infty}^{\infty}\mathrm{d}q_x\sum_{q_y}V(q)
 \left[ L_N\left(\frac12 q^2\ell^2\right) \right]^2 \\
 &\times \mathrm{e}^{
 	-\frac12 q^2\ell^2+iq_x(j_1-j_3)\frac{2\pi\ell^2}{L_y}
 	} \delta_{q_y,\frac{2\pi (j_1-j_4)}{L_y}}\delta_{j_1+j_2,j_3+j_4}
 \end{align}
 where $L_N(z)$ is Laguerre polynomial and $q^2 = q_x^2+q_y^2$. $V(q)$ is the Fourier transformation of the electron-electron interation:
 \begin{equation}
 V(q) =
 \left\{
 \begin{aligned}
 &\sum_m v_m L_m(q^2) \\
 &\frac{e^2}{\epsilon}\frac{\xi}{2} \sqrt{\pi} \exp{\left(-\frac{q^2\xi^2}{8}\right)}
 I_0\left(\frac{q^2\xi^2}{8}\right)
 \end{aligned}\right.
 \end{equation}
 The first line describes Haldane pseudopotential~\cite{Haldane1983}. The second line is the result of modified Coulomb interaction
 and $I_0(z)$ is the first kind of modified Bessel function. The factor $\delta_{j_1+j_2,j_3+j_4}$ in Eq. (\ref{coef}) indicates
 momentum conservation $j_1+j_2=j_3+j_4$.

 \subsection{Operator ordering and Fermionic phase}
 For the convenience of calculation in DMRG, we should rearrange the operator ordering in Eq.(\ref{oriH}) and transform the fermionic
 operators into hard-core bonsonic operators. In the sweep process of DMRG, we need calculate the left(right) environment Hamiltonian from
 left(right) to right(left), thus the ordering of operator string of some local operators must be ascending ordering from left to right,
  i.e. the operator acts on the left site to the left of string.
 For example, $\hat c^\dagger_1\hat c_6\hat c_4\hat c^\dagger_9$ is invalid but $-\hat c^\dagger_1\hat c_4\hat c_6\hat c^\dagger_9$ is valid. By imposing the ascending ordering
 to Eq. (\ref{oriH})
  \begin{align}\label{HI_DMRG}\nonumber
 \hat H_I = & \sum_{j_1<j_2,j_3<j_4}\mathcal{A}^N_{j_1,j_2,j_3,j_4}
 \hat c^\dagger_{j_1}\hat c^\dagger_{j_2}\hat c_{j_3}\hat c_{j_4} \\ \nonumber
 =& \quad \sum_{j_1<j_3<j_4<j_2}\mathcal{A}^N_{j_1,j_2,j_3,j_4}
 \hat c^\dagger_{j_1}\hat c_{j_3}\hat c_{j_4}\hat c^\dagger_{j_2}
 \left(
 \overset{\curvearrowleft}{\underset{j_1}{\circ} ~ \underset{j_3}{\bullet}}
 ~ \overset{\curvearrowright}{\underset{j_4}{\bullet} ~ \underset{j_2}{\circ}}
 \right) \\ \nonumber
 &+ \sum_{j_3<j_1<j_2<j_4}\mathcal{A}^N_{j_1,j_2,j_3,j_4}
 \hat c_{j_3}\hat c^\dagger_{j_1}\hat c^\dagger_{j_2}\hat c_{j_4}
 \left(
 \overset{\curvearrowright}{\underset{j_3}{\bullet} ~ \underset{j_1}{\circ}}
 ~ \overset{\curvearrowleft}{\underset{j_2}{\circ} ~\underset{j_4}{\bullet}}
 \right) \\
 &- \sum_{j_1=j_3<j_2=j_4}\mathcal{A}^N_{j_1,j_2,j_1,j_2} \hat n_{j_1}\hat n_{j_2}
 \end{align}
 where $\mathcal{A}^N_{j_1,j_2,j_3,j_4}=\left(2A^N_{j_1,j_2,j_3,j_4}-2A^N_{j_2,j_1,j_3,j_4}\right)$.

 Finally, we use the Jordan-Wigner transformation to transform fermion to hard core boson
 \begin{align}\label{HI_DMRG_F}\nonumber
 H_I
 =& - \sum_{j_1<j_3<j_4<j_2}\mathcal{A}^N_{j_1,j_2,j_3,j_4}
 a^\dagger_{j_1}F_{j_1,j_4}\hat a_{j_3}\hat a_{j_4}\hat F_{j_4,j_2}\hat a^\dagger_{j_2} \\ \nonumber
 &- \sum_{j_3<j_1<j_2<j_4}\mathcal{A}^N_{j_1,j_2,j_3,j_4}
 \hat a_{j_3}\hat F_{j_3,j_1}\hat a^\dagger_{j_1}\hat a^\dagger_{j_2}\hat F_{j_2,j_4}\hat a_{j_4} \\
 &- \sum_{j_1=j_3<j_2=j_4}\mathcal{A}^N_{j_1,j_2,j_1,j_2} \hat n_{j_1}\hat n_{j_2}
 \end{align}
 where $\hat F_{i,j}=\Pi_{s=i+1}^{j-1}\hat F_s=(-1)^{\sum_{s=i+1}^{j-1} \hat n_s}$ is the Jordan-Wigner string operator (fermionic phase operator)
 and $\hat a^\dagger(\hat a)$ is hard core boson operator. All operators in Eq(\ref{HI_DMRG_F}) are local and all operator strings are ascending
 ordering from left to right. Conclusively, Eq(\ref{HI_DMRG_F}) is a DMRG-friendly form of many-body Hamiltonian Eq(\ref{oriH}).

 \subsection{Conserved quantities}
 Generally speaking, the invariance of Hamiltonian under symmetry transformation gives us conserved quantum number. First we need to find all symmetry groups which Hamiltonian is invariant under the symmetric operations. Since the generators of different group may not commute with each other, we can only use the symmetry group which their generators are commute with each other. These generators can be diagonalized with Hamiltonian simultaneously. The eigenvalues of these generators are good quantum numbers of Hamiltonian.

 First, under a global gauge transformation of Landau level $\psi_{N,j}\rightarrow\mathrm{e}^{i\theta}\psi_{N,j}$, which is $U(1)$ group,
 the Hamiltonian is invariant under this global gauge transformation. The conversed quantity corresponds to this symmetry is particle
 number $[\hat N_{e}, \hat H]=0$, where
  \begin{equation}
  \hat N_{e}=\sum_{j=1}^{N_\phi}\hat n_j=\sum_{j=1}^{N_\phi}\hat c_j^\dagger \hat c_j.
  \end{equation}

 Second, we consider the translational operator $\hat T(\bm{a})$~\cite{Bernevig2012}
  \begin{equation}\label{TO}
  \hat T(\bm a) = \prod_{i=1}^{N_e} \hat T_i(\bm a) = \prod_{i=1}^{N_e}
  \exp{\left( \frac{i}{\hbar} \bm a \cdot  \hat{\bm K}_i \right) }
  \end{equation}
  where $\hat{\bm K}_i=\hat{\bm \Pi}_i-|e|\bm{B}\times\hat{\bm{r}}_i$($\bm{e}$ is unit vector and the subscript $i$ is single particle label) is called guiding center
  momentum and $\hat{\bm \Pi}_i=\hat{\bm p}_i+|e|\bm{A}$ is canonical momentum.
  $\hat{K}_{i,y} = \hat{p}_{i,y}$ and $\hat{K}_{i,x} = \hat{p}_{i,x}-\frac{\hbar}{\ell^2}\hat{y}_i$,
  where the second subscript denotes the component of vector.
  One can prove that $[\hat T(\bm a), \hat H]=0$ \cite{Haldane1985,Bernevig2012}.
  Crucially, the translation operators along different directions do not commute with each other,
  it can be seen from the following relation
  \begin{equation}\label{guiding-center-commutation}
  [\hat{K}_{i,x},\hat{K}_{j,y}] = i\hbar |e|B_z\delta_{ij}
  = -i\frac{\hbar^2}{\ell^2}\delta_{ij}.
  \end{equation}
  Since our setup of cylinder is periodic along $y$-direction, we have
  $\hat T(a\bm{e}_y) \psi_{N,j}(x,y)=\psi_{N,j}(x,y+a) = \mathrm{e}^{i\frac{2\pi j}{L_y}a}\psi_{N,j}(x,y) $, which leads to
  the second conserved operator  $\hat{K}_y=\sum_{i=1}^{N_e}\hat{K}_{i,y}$:
  \begin{equation}
  \hat{K}_y |j_1\cdots j_{N_e}\rangle = \frac{2\pi \hbar}{L_y}\sum_{i=1}^{N_e} j_i|j_1\cdots j_{N_e}\rangle
  \end{equation}
  Thus, the conserved quantity is
  \begin{equation}\label{momentumF}
  	\langle \hat{K}_y \rangle = \frac{2\pi \hbar}{L_y}\sum_{i=1}^{N_e} j_i.
  \end{equation}
  There is another definition of momentum
  \begin{equation}\label{momentumI}
  \langle \frac{L_y}{2\pi \hbar}\hat{K}_y \rangle = \sum_{i=1}^{N_e} j_i.
  \end{equation}
which is convenient in DMRG routine.
These two definitions only differ by a constant.
Unless otherwise specified, we choose the second definition in this paper.

 Finally, we consider how the operator $\hat T(a\bm{e}_x)$ affects the momentum $\hat{K}_y$.
 The special translation distances $X_m=\frac{2\pi \ell^2}{L_y}m (m\in\mathbb{Z})$ along $x$-direction which translate one single orbital to another are important in DMRG agorithm, and these operators $\hat{T}_{m}=\hat T(-X_m\bm{e}_x)$ act on single orbit
 \begin{equation}
 	\hat{T}_{m}\psi_{N,j}(x,y)=\psi_{N,j}(x-X_m,y)=\psi_{N,j+m}(x,y) ,
 \end{equation}
 the momentum of this single orbit changes from $j$ to $j+m$, which can also be seen by considering
 \begin{equation}\label{momentumchange}
 	\hat{T}^\dagger_m\hat{K}_y\hat{T}_m = \hat{K}_y + \frac{2\pi\hbar}{L_y}m\hat{N}_{e}.
 \end{equation}
 Eq.(\ref{momentumchange}) is an important relation in some situations of DMRG agorithm.

\subsection{DMRG algorithm}\label{sec::DMRG}
Due to the complexity of FQH systems, the exact diagonalization (ED) research is too difficult to extend to large system sizes.
In this work, to reach large-sizes and discuss the edge and interface physics, we will rely on Density-matrix renormalization group (DMRG) method.
DMRG algorithm ~\cite{White1992,Schollwock2011} is a powerful technique for studying 1D systems. For the short-range correlation length of ground states of gapped system, one can truncate the Hilbert space to small dimension with high accuracy. In this section, we will illustrate the implementation of DMRG algorithm in FQH system \cite{DMRGFQH2008,JizeZhao2011,Zaletel2013,Zaletel2015}. All algorithms in this paper are implemented based on the ITensor library(C++ 3)\cite{itensor}.

\subsubsection{Finite DMRG}
Finite DMRG (fDMRG) is a variational method to find the ground state of a system on a given finite size.
First we give a initial state in form of matrix product state (MPS), the goal is to find the ground state of Eq.(\ref{HI_DMRG_F}).
Since the incompressible natural of FQH state, the correlation and entanglement is finite for ground state. Thus, we
can truncate MPS by diagonalizing the reduced density matrix(or singular value decomposition for MPS, i.e SVD) and keep the largest $D$ eigenstates and eigenvalues(or singular values, the square root of eigenvalues), where $D$ is called the bond-dimension of MPS and the summation of all discarded eigenvalues is called truncation error $\epsilon$. In the limit $D\rightarrow\infty$, the truncation error becomes zero and the MPS representation becomes exact.
And we will discuss the convergence of  fDMRG in Sec.\ref{convergence}.

Since the Hamiltonian Eq.(\ref{HI_DMRG_F}) commutes with $\hat T(a\bm{e}_y)$, the momentum of the initial state is invariant in fDMRG,
so we can only reach a ground state with the same momentum as the initial state. This provides a way to select or control the ground state, by starting from an initial state with specific momentum.
For instance, for Laughlin $\nu=1/3$ state, the  root configuration is $|010010\cdots\rangle$, and the model ground state is superposition of
 $|010010\cdots\rangle$ and its “squeeze” sequences ~\cite{HaldaneJack2008}.
Note that the third line of Eq.(\ref{HI_DMRG}) actually describes a typical “squeeze” and therefore we have possibility to reach the ground state in fDMRG.

\subsubsection{Improvement of DMRG}\label{convergence}
In this subsection, we will discuss some techniques of DMRG in solving the ground state of Eq.(\ref{HI_DMRG_F}). These tricks are important for DMRG in the FQH systems, or DMRG in the momentum space. To proceed, let us briefly explain why we highlight these techniques.
Let us consider the most simple Hamiltonian with only the nearest neighbour interaction (e.g. $\hat{\bm{S}}_i\cdot\hat{\bm{S}}_{i+1}, \hat{c}^\dagger_i\hat{c}_{i+1}$), the two-site DMRG can find a optimal state even if the initial state is a product state. The  two-site DMRG can only ``see" the nearest neighbour interaction, or say it can only find local
optimal wavefunction in the two-site Hilbert space, for example $\mathcal{H}_2 = \text{span} (|00\rangle, |10\rangle, |01\rangle, |11\rangle)$. Once we give a product state like $|010010\cdots\rangle$, the momentum conservation prohibits the nearest neighbor hopping,  hence the initial state $|010010\cdots\rangle$ never change and DMRG algorithm will fall into a local solution forever.

Due to this reason, here we list some methods to overcome this drawback of two-site DMRG as below.

\textit{n-site DMRG}: Since the momentum conservation prohibits the nearest neighbor hopping and the simplest “squeeze” step($|1001\rangle\rightarrow|0110)\rangle$) need at least 4-site optimazation. The most straightforward way is to use n-site optimization instead of 2-site.
 	The n-site algorithm for some $n>2$ depends on the filling ~\cite{Zaletel2013}.
 	This significantly increases the memory requirements of the algorithm.
 	
 	\textit{Superposition initial state}: The 2-site update can not reach optimal state from initial state
 	$|010010\cdots\rangle$. Since the model state is superposition of root configuration and its
 	“squeeze”, therefore we can use the superposition of a set of “squeezed” states instead of a
 	product state~\cite{DMRGFQH2008}.
 	
 	\textit{Global subspace expansion}: Global subspace expansion(GSE) is proposed by
 	Ref.~\cite{GSETDVP2020}, which is used to improve the traditional time-dependent variantional principle(TDVP) agorithrm and avoid falling into local solutions. The key idea of
 	GSE is the enlargement of the tangent space by Krylov subspace of order $k$
 	\begin{equation}
 		\mathcal{K}(\hat H,|\psi\rangle) = \text{span}\{ |\psi\rangle,
 		\hat{H} |\psi\rangle, \cdots \hat{H}^{k-1} |\psi\rangle\},
 	\end{equation}
 	and use the state in this Krylov subspace instead of $|\psi\rangle$.
 	
 	\textit{Density matrix correction}: Density matrix correction  \cite{DenMatCorWhite}  can improve the convergence dramatically. The main idea
 	 is illustrated as follow. Eq.(\ref{HI_DMRG_F}) can be written as $\hat H = \sum_i
 	 \hat{H}_{L,i}\otimes\hat{H}_{R,i}$ and $|\psi\rangle = \sum_i s_i |L_i\rangle\otimes|R_i\rangle$. The reduced density matrix correction is add the corrected term
 	 $\Delta\rho=\sum_i \langle R_i |\psi'\rangle\langle \psi'|R_i\rangle $
 	 to original reduced density matrix
 	 $\rho=\sum_i \langle R_i |\psi\rangle\langle \psi|R_i\rangle  $
 	 where
 	 $ |\psi'\rangle = \sum_{i} \hat{H}_{L,i}\otimes\hat{I}_R|\psi\rangle $.
 	 The final step is to replace the original reduced density matrix $\rho$ with the new one
 	  $\rho+\delta\Delta\rho$, where $\delta$ is a small controllable weight constant.
 	  Sometimes $\Delta\rho$ is regarded as a “noise” term, then $\delta$ is the intensity of “noise”.

Lastly, when applying finite DMRG to the cylinder geometry,  to avoid the electrons trapped at the two ends of the finite cylinder, it is necessary to include an additional trap  potential \cite{RotonEdge}. The selection of trap potential is usually empirical. An alternative way to overcome this issue is to work on the infinite cylinder geometry, as discussed below.

\subsubsection{Infinity DMRG}

\begin{figure}[t]
 	\includegraphics[width=0.46\textwidth]{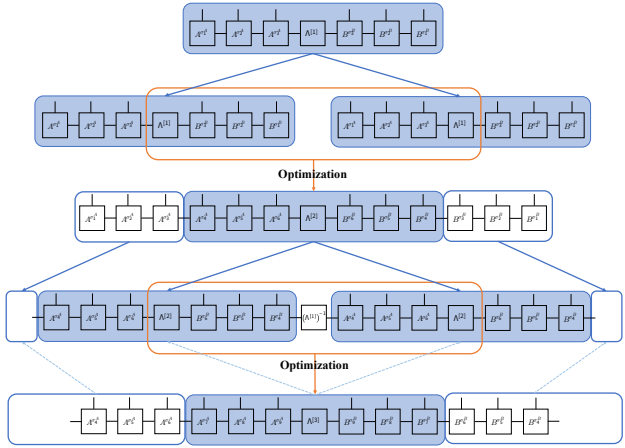}
 	\caption{\textit{The main steps of calculating Laughlin $\nu=1/3$ state using iDMRG:}
 		(First line) Starting from an unit cell with $2q=6$ orbits and optimizing the energy of MPS.
 		(Second and third line) Then we enlarge the MPS by inserting a new unit cell and optimize the energy of this new unit cell, here we have used the initial guess.
 		(Fourth and fifth line) By repeating enlargement and optimization step, we can reach the thermodynamic limit and derive a translational invariant unit cell.
 	}\label{fig:iDMRG}
 \end{figure}

Infinity-DMRG (iDMRG) is a powerful method for the FQH systems  \cite{Zaletel2013}. In this subsection, we will discuss the technical details for implementing iDMRG
to FQH systems. We choose Laughlin $\nu=1/3$ state as an example, and show the main process of iDRMG in Fig. \ref{fig:iDMRG}.

\textit{Step 1}: Starting from an unit cell with $2q=6$ orbits, using fDMRG method to minimize the energy, then we have the ground state MPS:
\begin{equation}\label{iDMRG:step1}
|\psi_1\rangle = \sum_{\sigma} A^{\sigma_1^A}A^{\sigma_2^A}A^{\sigma_3^A} \Lambda^{[1]} B^{\sigma_3^B}B^{\sigma_2^B}B^{\sigma_1^B} | \sigma \rangle .
\end{equation}
Where $A^\sigma$ and $B^\sigma$ are left-orthogonal and right-orthogonal tensors respectively.

\textit{Step 2:}After the first step, we insert a new unit cell into the center of the original MPS. Following the intuition of MPS, we write down the ``enlarged" MPS as:
\begin{equation}
|\psi_2\rangle = \sum_{\sigma} L
\left(A^{\sigma_4^A}A^{\sigma_5^A}A^{\sigma_6^A} \Lambda^{[2]} B^{\sigma_6^B}B^{\sigma_5^B}B^{\sigma_4^B} \right) R | \sigma \rangle
\end{equation}
where $ L = A^{\sigma_1^A}A^{\sigma_2^A}A^{\sigma_3^A} $
and $R  = B^{\sigma_3^B}B^{\sigma_2^B}B^{\sigma_1^B}$.
The crucial task is to find a good initial guess for the newly inserted tensors which can help to
quickly reach the translational invariant state in the calculation.
Following the key ideas in Ref.\cite{Schollwock2011}, the good
initial guess is to translate the right half(including the tensor containing singular singular values) of Eq.(\ref{iDMRG:step1}) to the left half of the newly
inserted unit cell, and vice verce for the other half. Therefore, the initial guess can be written as
\begin{equation}
|\psi_2\rangle = \sum_{\sigma} L
\left(\Lambda^{[1]} B^{\sigma_3^B}B^{\sigma_2^B}B^{\sigma_1^B} A^{\sigma_1^A}A^{\sigma_2^A}A^{\sigma_3^A} \Lambda^{[1]}\right)R | \sigma \rangle .
\end{equation}
It is worth noting that we have translated the tesnors in this step, the conserved quantities of each tensors need to be treated carefully.
Since the generators of translation operators are not commute with each other, translating tensors in MPS will change the quantum numbers
they carried. Following Eq.(\ref{momentumchange}), when we translate $A(B)$ and $\Lambda$ from position $i$ to $j$, the quantum numbers must
change from $(N_e, K)$ to $(N_e, K + (j-i)N_e)$.

\textit{Step 3}: Optimizing the newly inserted unit cell by fDMRG and keep the other tensors unchanged, we get the following MPS with 12 orbits:
\begin{equation}
|\psi_2\rangle = \sum_{\sigma} L
\left(A^{\sigma_4^A}A^{\sigma_5^A}A^{\sigma_6^A} \Lambda^{[2]} B^{\sigma_6^B}B^{\sigma_5^B}B^{\sigma_4^B} \right)R | \sigma \rangle .
\end{equation}

\textit{Step 4}: Using the same method to insert a new unit cell into the middle again
\begin{equation}
|\psi_3\rangle = \sum_{\sigma} L
\left(\Lambda^{[2]} B^{\sigma_6^B}B^{\sigma_5^B}B^{\sigma_4^B} (\Lambda^{[1]})^{-1} A^{\sigma_4^A}A^{\sigma_5^A}A^{\sigma_6^A}\Lambda^{[2]}  \right)R | \sigma \rangle
\end{equation}
where $L = A^{\sigma_1^A}A^{\sigma_2^A}A^{\sigma_3^A} A^{\sigma_4^A}A^{\sigma_5^A}A^{\sigma_6^A}$
and $R = B^{\sigma_6^B}B^{\sigma_5^B}B^{\sigma_4^B}B^{\sigma_3^B}B^{\sigma_2^B}B^{\sigma_1^B}$.
By repeating \textit{Step 3} and \textit{Step 4}, we can reach the thermodynamic limit and derive a translational invariant unit cell.

\subsection{The construction of interface}\label{cons_interface}

\begin{figure}[t]
	\includegraphics[width=0.46\textwidth]{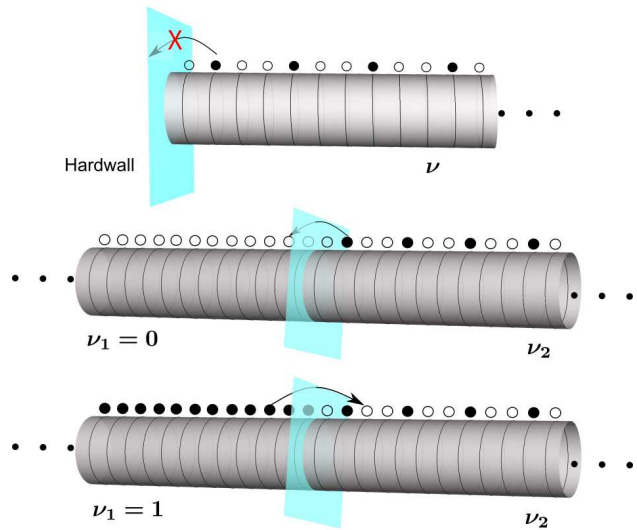}
	\caption{\textit{The construction of interface:}
		(Top) The schematic diagram of constructing the ``hardwall edge", where just drop the left boundary.
		(Middle and bottom) The construction of the interface between two different QH states. We first cut each of the two infinte cylinders into two halves, and glue the left half of $\nu_1$ state and the right half of $\nu_2$ state together. The blue plane denotes the interface. 
	}\label{fig::DMRGinterface}
\end{figure}

In this paper, we will investigate the interface physics of FQH states. This is partially motivated by the recent experimental works (e.g. Ref. \cite{NovelDutta2021}).
Here we explain the ``cut-and-glue" scheme that we construct the interface of FQH states. The general steps are as follows: 1) After the iDMRG algorithm converges, we get the translation-invariant MPS unit cell, we cut the MPS into two halves, i.e. the left part and the right part. 2) We drop the one part (say the left one) and project the bond between the left boundary and the first orbit of the right part, to the state has the same quantum number as the root configuration. 3) Now we have a semi-infinite cylinder with a filling factor $\nu_1$, and then we glue it with another semi-infinite cylinder with filling factor $\nu_2$ together, see Fig. \ref{fig::DMRGinterface}. 4) We optimize the energy of MPS around the interface region and find the lowest energy configuration, as the usual finite DMRG procedure.

In this paper we will construct three types of interfaces. (we will use ``$\nu_1|\nu_2$ interface" to denote the interface between $\nu_1$ state and $\nu_2$ state.)
\begin{itemize}
	\item The first one is called ``hardwall edge" (or with a sharp cleaved edge)\cite{Wan2003}, where the electron cannot enter the hardwall, or it can be seen as a situation where the other half takes infinite (sharp) confining potential (see Fig. \ref{fig::DMRGinterface} (top)). In this case, after we cut the translation-invariant MPS into two-halves, we can directly optimizes the semi-infinite cylinder by skipping the glue step.
    \item The second type is to glue a integer quantum Hall state with filling $\nu_1$ and the FQH  state with filling $\nu_2$ together, where $\nu_1$ is smaller than $\nu_2$  (see Fig. \ref{fig::DMRGinterface} (middle)). For example, $\nu_1=0|\nu_2=2/3$ or $\nu_1=2|\nu_2=8/3$ interface, the integer quantum Hall half is used to gap out the integer edge modes of FQH half. Please note that, in this case the electron could enter the integer filling region, in contrast to the ``hardwall edge".
    \item The third type is the similar to the second, but $\nu_1$ is greater than $\nu_2$  (see Fig. \ref{fig::DMRGinterface} (bottom)), for example, $\nu_1=1|\nu_2=2/3$ or $\nu_1=3| \nu_2=8/3$. In this case, tunneling through the interface is allowed.
\end{itemize}

Finally, we stress this construction scheme has many advantages.
First, the construction is based on the iDMRG scheme on the cylinder geometry \cite{Zaletel2013}, so that we can get a target FQH  state in an unbiased way, without any empirical information like confining potentials or the background charge layer.
Second,  after gluing step we perform a energetically variational optimization, so the obtained configurations (shown below) on the interface are the lowest energy configuration.
Next, this scheme will be displayed to investigate edges and interfaces in greater detail.

\subsection{Topological characters in the bulk and on the interface}\label{sec::topo}

In this section, we summarize the physical quantities that we will utilize for identifying the topological properties in the bulk and on the interface. They include entanglement spectra for chiral edge state, topological entanglement entropy for quasiparticle quantum dimension, momentum polarization for topological spin and guiding center spin, and electric dipole momentum for guiding-center spin.

\subsubsection{Entanglement spectra}
FQH states are gapped in the bulk but contain gapless edge excitations. It means that the
electrons in bulk are localized, and the transport properties(electrical and thermal conductance)~\cite{QuasiPartiNature,QuasiPartiPRL,5/2QuasiPartiNature, 5/2QuasiPartiScience,NeutralNature,5/2ThermalNature} are determined by the gapless edge states which are extended. This implies that there is a certain connection between edge and bulk states. On the other word, the topological order of a FQH state can be determined by its edge states~\cite{Wen1992}. Moreover, the entanglement spetrum of a FQH groundstate wavefunction can be seen as some special ``energy levels" which correspond to the entanglement Hamiltonian $\hat{\mathcal{H}}_E$.
It has been shown\cite{QiXL2012,Haldane2008} that there is a general relationship between the entanglement spectrum and the edge state spectrum of topological quantum states. Thus, one can investigate the entanglement spectra of a FQH state instead of the true edge excitations, provides a straightforward way to study the edge excitations. We briefly describe the method of calculating the entanglement spectra below.

Once we get the ground state wavefunction $|\psi\rangle$, consider a bipartition of the system into two part A and B, and the Hilbert space $H=H_A\otimes H_B$. One can write $|\psi\rangle$ as a superposition of some product states $|\psi\rangle = \sum_{i,j}c_{i,j}|\psi_i^A\rangle\otimes|\psi_j^B\rangle$ where $|\psi_i^A\rangle\in H_A$ and $|\psi_j^B\rangle\in H_B$. By employing the Schimidt decomposition(SVD), $|\psi\rangle$ has a special form
 \begin{equation}
 	|\psi\rangle = \sum_{i} \lambda_i |\psi_i^A\rangle\otimes|\psi_i^B\rangle
 \end{equation}
where $\lambda_i = \exp{(-\frac12 \xi_i)}\ge 0$ and $\xi_i$'s can be seen as ``energy levels"~\cite{Haldane2008}.
One can write the reduced density matrix as $\hat{\rho}_A = \mathrm{tr}_B |\psi\rangle\langle\psi| = \sum_i\exp{(-\xi_i)}|\psi_i^A\rangle\langle\psi_i^A|$ which is diagonal, thus we can see $\xi_i$ as engenvalues of $\hat{\mathcal{H}}_E = -\ln(\hat{\rho}_A) = \sum_i \xi_i|\psi_i^A\rangle\langle\psi_i^A|$ where $\hat{\mathcal{H}}_E$ is called ``entanglement Hamiltonian" and $\xi_i$'s are ``entanglement spectra" (ES). Therefore we can calculate the ES through $\xi_i = -\ln{(\lambda_i^2)}$.

The remarkable feature of ES is the low energy excitations. For example, the $U(1)$ free chiral boson Hamiltonian $H_e = \sum_{k>0}\epsilon(k)b^\dagger_kb_k$ is the effective Hamiltonian of Laughlin state, and the counting of low energy excitations obey~\cite{Wen1992}
 \begin{align}\nonumber
 		\Delta k&: \quad 0\quad 1\quad 2\quad 3\quad 4\quad 5\quad 6 ~~\cdots\\\nonumber
\text{Degeneracy}&:\quad 1\quad 1\quad 2\quad 3\quad 5\quad 7\quad 11\cdots
 \end{align}
Alternatively, the typical  counting ``1,1,2,3,5,7..." in the ES   tells us the effective edge theory is described by $U(1)$ free chiral boson theory.

\subsubsection{Area law and topological entanglement entropy}
A prominent measure of the entanglement between one part and the rest of a quantum many-body system is entanglement entropy(EE)~\cite{EE2005}. The Neumann entropy $S_E=-\mathrm{tr}(\hat{\rho}_A\ln \hat{\rho}_A)$ of reduced density matrix $\hat{\rho}_A=\sum_i\exp{(-\xi_i)}|\psi_i^A\rangle\langle\psi_i^A|$ is
\begin{equation}
	S_E=-\sum_i \lambda_i^2 \ln\lambda_i^2 = \sum_i \xi_i\exp{(-\xi_i)}.
\end{equation}
Generally, for a gapped state, the EE $S_E$ obeys ``area law" which means the EE is proportional to the ``area" of the joint boundary. Importantly, in addition to the area-law, there is an emergence of a nonzero constant term for the topologically ordered states, called topological entanglement entropy(TEE)~\cite{Kitaev2006,LevinWen2006}.

Now, we focus on the FQH states on a cylinder. When we cut the cylinder along the periodic direction, the area of the joint boundary is the circumference $L_y=L$ and the EE of a FQH state scales as~\cite{Lauchli2010}
\begin{equation}
	S_E = \alpha L - \gamma + \mathcal{O}(1/L)
\end{equation}
where $\gamma$ is TEE and $\alpha$ is a nonuniversal constant. The TEE $\gamma$ is a topological quantity which can be used to determine the topological order, $\gamma$ has a theoretical value $\gamma_a = \ln(\mathcal{D}/d_a)$, where $d_a$ is the quantum dimension of anyon a and $\mathcal{D}=\sqrt{\sum_a d_a^2}$~\cite{LevinWen2006} is the total quantum dimension of its topological field theory.

\subsubsection{Momentum polarization}
Following Ref.\cite{MomentPolar}, the topologically ordered state on a cylinder corresponds to the topological quasipaticle in each side. And the rotation along the periodic direction will not give us any information since the rotational invariance of cylinder. If one can rotate one side and keep another side unchanged, this progress will give us a phase which contains informations of the quasiparticle on this side. This idea can be realized by dividing the cylinder into two parts along the periodic direction. We denote these two parts as A and B and apply the operator $\hat{T}^A_y$ to A part, where $\hat{T}^A_y$ is the restriction of Eq.(\ref{TO})
\begin{equation}
	\hat{T}^A_y = \prod_{i\in A}\exp{\left( \frac{i}{\hbar} L_y \hat{K}_{i,y} \right) }
\end{equation}
 we obtain the phase
 \begin{equation}
 \exp{(i2\pi\langle M\rangle_a)} = \langle \psi_a|\hat{T}^A_y|\psi_a\rangle = \mathrm{tr}_A\left(\hat{\rho}_A\hat{T}^A_y \right)
 \end{equation}
 where $|\psi_a\rangle$ denotes the groundstate corresponds to topological sector $a$. Since $\hat{T}^A_y$ commutes with Hamiltonian, therefore we can diagonalize $\hat{\rho}_A$ and $\hat{T}^A_y$ simultaneously. We can calculate $\langle M\rangle_a$ from ES by
 \begin{equation}\label{MP}
 	\langle M\rangle_a = \mathrm{tr}_A\left(\hat{\rho}_A\frac{L_y}{2\pi\hbar}\hat{K}_{A,y} \right)
 	=\sum_i \mathrm{e}^{-\xi_i}\langle\frac{L_y}{2\pi\hbar}\hat{K}_{A,y}\rangle_i
 \end{equation}
 where $\hat{K}_{A,y} = \sum_{i\in A}\hat{K}_{i,y}$ is the guiding center momentum operator in A part and $\langle\hat{\mathcal{O}}\rangle_i = \langle\psi_i^A|\hat{\mathcal{O}}|\psi_i^A\rangle$.
 This phase $\langle M\rangle_a$ contains three important topological quantities
 \begin{equation}\label{EQ::MP}
 	\langle M\rangle_a-\langle M\rangle_{root} = \frac{\eta^g}{2\pi\hbar}L_y^2-h_a+\frac{c-\nu}{24}
 \end{equation}
 where $\eta^g=-\frac{\hbar}{4\pi\ell^2}\frac{s}{q}$ is guiding center Hall viscosity~\cite{Avron1995} and $s$ is guiding center spin~\cite{HaldaneGeometric2011,Read2009,Read2011,YJPark2014}. The subscript ``root" in $\langle M\rangle_{root}$ means that this term is calculated using the root state $|\psi_{root}\rangle$~\cite{YJPark2014}. The second term $h_a$ is the topological spin correspond to quasiparticle $a$~\cite{Zaletel2013, MomentPolar} and $c$ is the central charge of the underlying edge conformal field theory(CFT). The last term $\nu=p/q$ is the filling factor of a FQH state. These three quantities $s,h_a,c$ provide rich topological information, first $s$ is a quantized quantity correspond to the nondissipative response of the metric perturbation~\cite{TwistED}, similar to that the Hall conductance is the response of electromagnetic gauge $\bm{A}$. The other two quantities $h_a$ and $c$ are the elements of modular-$\mathcal{T}$ matrix, which is the unitary transformation of the groundstate manifold under modular transformation. Moreover, modular matrix can be used to describe the topological order~\cite{WenXGModular1993}.

\subsubsection{Electric dipole moment and guiding center spin}
Near the edge or around the interface, the intrinsic or external electric field inevitably induces a nonuniform density distribution of electrons. One typical result is the chiral edge mode forms and propagates along the edge or interface. Recently, it has been realized that, the nonuniform density distribution gives a quantized dipole moment, which is connected to the finite Hall viscosity of FQH liquids \cite{YJPark2014}, and it has been numerically verified in several systems \cite{YJPark2014,ZhuW2020}.

To be specific, the intrinsic dipole moment coupled to  the electric field generates an electric force. This electric force is to counter-balance the viscous force applied on the edge modes, which gives the relation \cite{YJPark2014}
\begin{equation}\label{vis_eletric_balance}
 \eta^g = \frac{p_y}{L_y}B_z = -\frac{p_y}{L_y}B
\end{equation}
where $B$ is magnetic field, $\eta^g=-\frac{\hbar}{4\pi\ell^2}\frac{s}{q}$ is guiding center Hall viscosity and $p_y$ is the dipole moment
\begin{equation}
 	p_y=-\frac{2\pi|e|}{L_y}\sum_i (\langle\hat{n}_i \rangle-\nu)i
\end{equation}
now we have
\begin{equation}\label{Eq::dipole}
 -\frac{s}{q} = \frac{8\pi^2}{L_y^2}\sum_i (\langle\hat{n}_i \rangle-\nu)i
\end{equation}
where $s$ is guiding center spin and $q$ is the denominator of $\nu=p/q$. In the following, we will show that the dipole moment is an important quantity, even though it is calculated by the nonuniversal oscillated density(denpends on the interaction), but the dipole moment(or guiding center spin) is protected by topology, for more details see Sec.\ref{sec::edge} and Ref\cite{YJPark2014}.

\subsubsection{Relations between guiding center spin and topological shift}
Generally, the total Hall viscosity of a FQH state includes two parts: 
\begin{equation}\label{Eq:viscosity}
\eta^H=\eta^o+\eta^g=\frac{\hbar}{4\pi\ell^2}\left( \nu\tilde{s}-\frac sq \right)
=\frac{\hbar\nu}{4\pi\ell^2}\bar s
=\frac{\hbar\nu}{8\pi\ell^2}\mathcal{S}
\end{equation}
Where $\eta^g=-\frac{\hbar}{4\pi\ell^2}\frac{s}{q}$ is the guiding center Hall viscosity and $\eta_o=\frac{\hbar}{4\pi\ell^2} \nu\tilde{s}$ is the Landau orbital Hall viscosity\cite{YJPark2014,TwistED}. The quantity $\tilde{s}=n+\frac12$ in $\eta_o$ is the Landau-orbital spin for the $n$th($n=0,1,2,\cdots$) Landau level.
 Since $\eta_o$ comes from the Landau-orbital form factor and we are working on the LL-projected Hamiltonian, thus it does not appear in Eq.(\ref{EQ::MP}) and Eq.(\ref{vis_eletric_balance}). Interestingly, by defining  the mean ``orbital spin" $\bar s=\tilde{s}-\frac sp$  \cite{Read2011}, one can  relate it to the topological shift $\mathcal{S}$\cite{Zee1992} via $\mathcal{S}=2\bar s$\cite{Read2011}. From
 Eq.(\ref{Eq:viscosity}), we can write down the relation between guiding center spin and topological shift
 \begin{equation}\label{shift}
 	\mathcal{S} = 2\left(\tilde{s}-\frac{s}{p} \right).
 \end{equation}

 Eq.(\ref{shift}) means that we can obtain topological shift by calculating guiding center spin since $\tilde{s}$ is a known number. Please note, the filling factor $\nu=p/q$ in both
  Eq.(\ref{Eq:viscosity}) and Eq.(\ref{shift}) is the filling of a single Landau level. For example, the $\nu=8/3$ state has total filling $\nu=2+2/3$, so its filling in the second Landau level is
  $\nu=2/3$ and we should use $\nu=p/q=2/3$ in Eq.(\ref{Eq:viscosity}) and Eq.(\ref{shift}). If the nature is belong to the Laughlin state, $\mathcal{S}$ is $0$ and $2$ for $2/3$ state and $2+2/3$ state, respectively. As a comparison, non-Abelian $Z_4$ Parafermion hosts $\mathcal S =3$\cite{Read1999,Barkeshli2010}, and Bonderson-Slingerland Hierarchy has $\mathcal S=4$\cite{Bonderson2008}(in LLL).
  Finally, we provide some details of the theoretical derivation of $\mathcal{S}$
   and $s$ for Abelian $\nu=n+1/q$ state and $\nu=n+1-1/q$ state in Appendix\ref{apdx::shift}.

 \begin{figure}[t]
 	\includegraphics[width=0.46\textwidth]{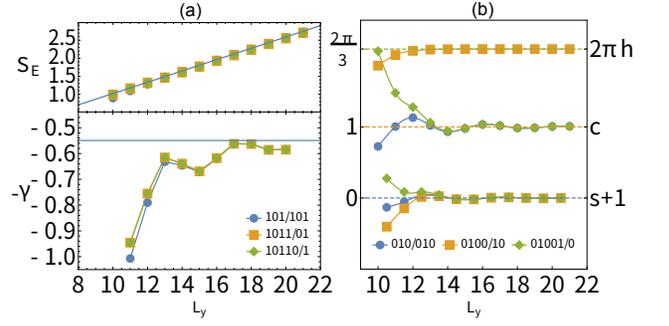}
 	\caption{\textit{The topological order of Laughlin $1/3$ state for Coulomb interaction:}
 		(a) The entanglement entropy of Laughlin $1/3$ state in different $L_y$. The upper panel is the scaling behavior of entanglement entropy $S_E\approx0.157L_y-\gamma$. The lower panel is the topological entanglement entropy $\gamma$ where the blue horizontal line is $-\gamma=-\ln{\sqrt{3}}$.
 		(b) The momentum polarization of Laughlin $1/3$ state in different $L_y$. The obtained topological charaters are guiding center spin $s\approx-1$, topological spin $h\approx\frac13$ and central charge $c\approx1$.
 	}\label{fig:1_3_Topological_Order_Cou}
 \end{figure}

 \begin{figure}[b]
 	\includegraphics[width=0.46\textwidth]{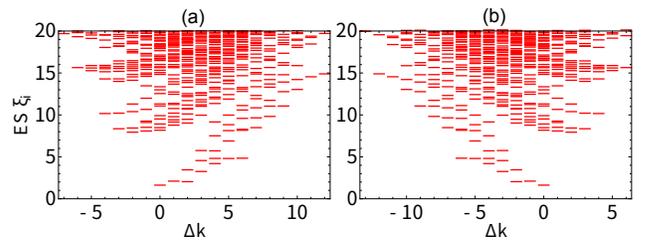}
 	\caption{\textit{The ES of $\nu=1/3,2/3$ state for Coulomb interaction:}
 		The entanlement spetrum of (a)Laughlin $1/3$ state and (b)$2/3$ state. The circumference is $L_y=20\ell$. Both two show the typical U(1) counting $1,1,2,3,5\dots$, but their chirality is opposite.
 	}\label{fig:ES_Cou_b_20}
 \end{figure}
\subsubsection{Relations between dipole moment and momentum polarization}

Now, we illustrate the relation between dipole moment and momentum polarization, which has also been explained in Ref\cite{YJPark2014}. Starting from Eq.(\ref{MP}), we have
$\langle M\rangle_a = \mathrm{tr}_A\left(\hat{\rho}_A\frac{L_y}{2\pi\hbar}\hat{K}_{A,y} \right)$,
 note that the momentum operator $\hat{K}_{A,y}$ can be expressed as
$\hat{K}_{A,y} = \sum_{i\in A}\hat{n}_i\frac{2\pi i\hbar}{L_y}$, using this relation we have
\begin{equation}
\langle M\rangle_a =\sum_{i\in A}\mathrm{tr}_A\left(\hat{\rho}_A\hat{n}_i \right)\times i
=\sum_{i\in A}\langle\hat{n}_i\rangle \times i.
\end{equation}
Comparing Eq.(\ref{EQ::MP}) and Eq.(\ref{Eq::dipole}), we can find that the Eq.(\ref{Eq::dipole}) should be
\begin{equation}\label{EQ::dipole_MP}
-\frac sq +\frac{8\pi^2}{L_y^2}\left(-h_a+\frac{c-\nu}{24} \right)
= \frac{8\pi^2}{L_y^2}\sum_i (\langle\hat{n}_i \rangle-\nu)i
\end{equation}
Here, we have found a correction on Eq.(\ref{Eq::dipole}) in order $O(L_y^{-2})$ which will vanish in thermaldynamical limit.  Please note the $O(L_y^{-2})$ coefficient on RHS Eq.(\ref{Eq::dipole}) can be absorbed by $\frac{8\pi^2}{L_y^2}\sum_i i \approx \int kdk $ where  $k=\frac{2\pi i}{L_y}$, and Eq.(\ref{EQ::dipole_MP}) can be written as
\begin{equation}
-\frac sq = \int dk (\langle\hat{n}(k) \rangle-\nu(k))k
\end{equation}
In Appendix\ref{apdx::modular}, we provide more details about the corrected term in Eq.(\ref{EQ::dipole_MP}).

\section{Topological properties  in the bulk} \label{sec::bulktopo}
In this section we will study the bulk properties of both $\nu=1/3,2/3$ state.
The nature of $\nu=2/3$ state is thought to be the particle-hole conjugation of Laughlin $\nu=1/3$ state~\cite{Girvin1984,MacDonald1990},
but the discussion on $\nu=2/3$ state revives recently \cite{Yinghai2012,Zixiang2008,Gefen2014}, mainly due to experimental identification complex edge structures that is beyond the previous description \cite{Bid2009,Cohen2019,NP2/3_2017}.
We also study the $\nu=2/3$ state in the second Landau level(SLL), i.e.
the $\nu=8/3$ state. It has been theoretically proposed this state could be non-Abelian.
These are the motivations for this work.

Here, we list the values of parameters we used in iDMRG.
(a)
We choose a cutoff $\epsilon_H=10^{-6}$ of Hamiltonian, which means we only consider terms with coefficients in Eq.(\ref{coef}) $ |A^N_{j_1,j_2,j_3,j_4}|\geq \epsilon_H$.
(b) The upper bound of truncation error in DMRG is chosen to be $\epsilon_D = 10^{-7}$ and the bond dimension of MPS is automatically increased to ensure the truncation error is smaller than $\epsilon_D$.
(c) The intensity of noise in density matrix correction is $\delta \approx 10^{-3}-10^{-4}$.
(d) The regulated length in Coulomb interaction is $\xi = 4\ell$. The full text will use the above setting of parameters.

\subsection{The Laughlin $1/3$ state}
The Laughlin $1/3$ state is the simplest FQH state and its topological characters have been extensively studied numerically and theoretically.
Here we use it as a benchmark.
The Laughlin $1/3$ state is an Abelian state with three kinds of quasiparticle $0, +|e|/3, -|e|/3$, and all of them have quantum dimension $d_a=1$ thus the total quantum dimension is $\mathcal{D} = \sqrt{3}$, so the TEE should be $\gamma_a = \ln{\sqrt{3}}\approx 0.549$.
The EE $S_E$ of Laughlin $1/3$ state obtained using iDMRG have been shown in Fig.\ref{fig:1_3_Topological_Order_Cou}(a).
Clearly the EE satisfies the area law: $S_E\approx 0.157 L_y-\gamma $. In the lower panel of Fig.\ref{fig:1_3_Topological_Order_Cou}(a), the TEE $\gamma$ converges to $\ln{\sqrt{3}}\approx0.549$ as $L_y$ increasing.

The guiding center spin of Laughlin $1/q$ state is $s = \frac{1-q}{2}$~\cite{TwistED} (or see Appendix\ref{apdx::shift}), so the theoretical value of Laughlin $1/3$ state is $s = -1$ and topological spin is $h=\frac13$~\cite{TwistED}. The numerical results of momentum polarization have been shown in Fig.\ref{fig:1_3_Topological_Order_Cou}(b), the results perfectly agreement with the theoretical prediction as $L_y$ increasing. Finally, the edge theory of Laughlin $1/3$ state is described by chiral free boson with central charge $c=1$. In Fig.\ref{fig:1_3_Topological_Order_Cou}(b), the obtained $c$ in momentum polarization perfectly converges to $1$. Meanwhile, the ES of Laughlin $1/3$ state have been shown in Fig.\ref{fig:ES_Cou_b_20}(a), it explicitly exhibits the chirality and the degeneracy of low energy excitations are $1,1,2,3,5\cdots$ which is consistent with theoretical prediction of chiral free boson.

 \begin{figure}[t]
 	\includegraphics[width=0.46\textwidth]{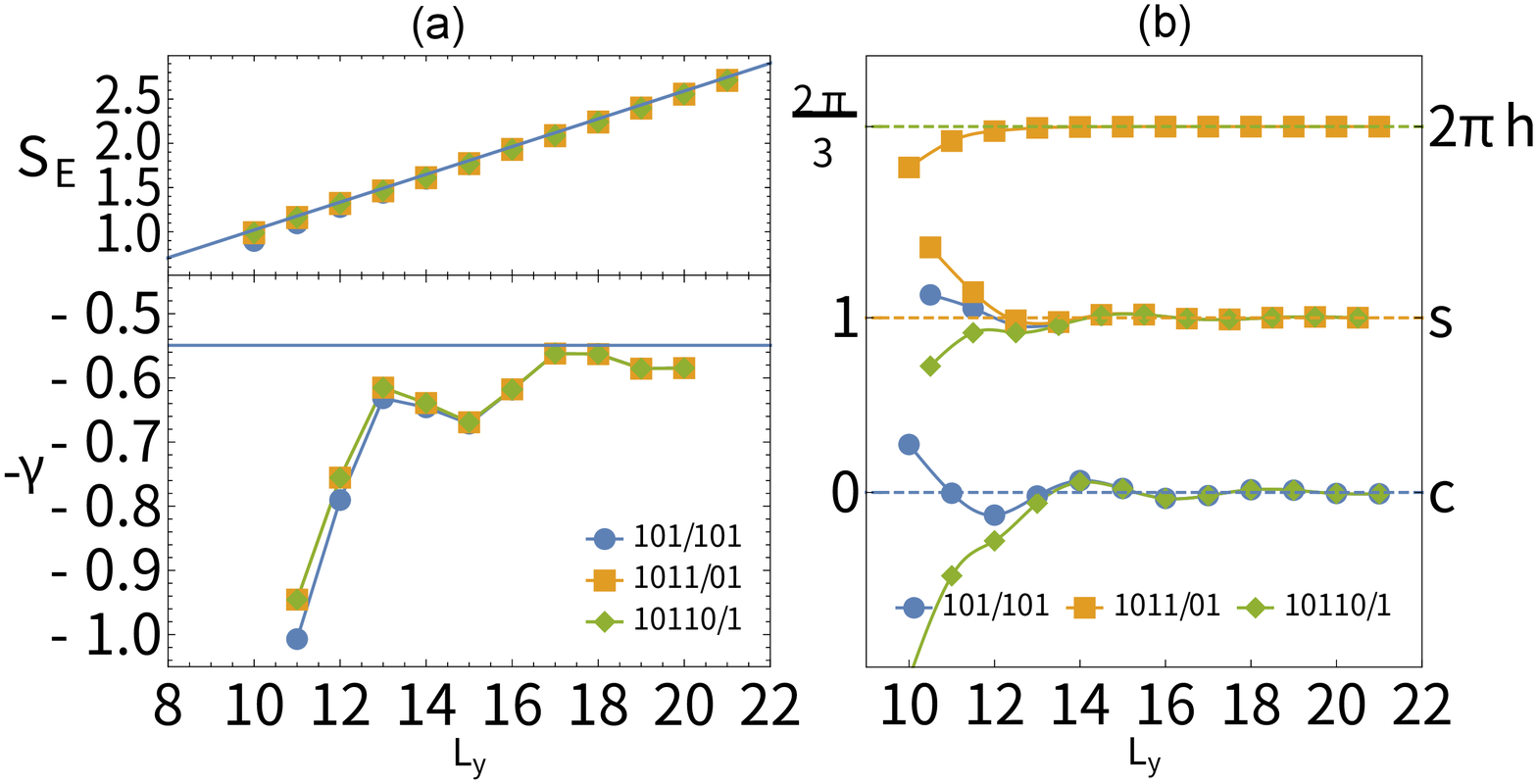}
 	\caption{\textit{The topological order of $2/3$ state for Coulomb interaction: }
 		(a) The entanglement entropy of $2/3$ state in different $L_y$. The upper panel is the scaling behavior of entanglement entropy $S_E\approx0.157L_y-\gamma$. The lower panel is the topological entanglement entropy $\gamma$ where the blue horizontal line is $-\gamma=-\ln{\sqrt{3}}$.
 		(b) The momentum polarization of $2/3$ state in different $L_y$. The obtained topological charaters are guiding center spin $s\approx1$, topological spin $h\approx\frac13$ and chiral central charge $c\approx0$.
 	}\label{fig:2_3_Topological_Order_Cou}
 \end{figure}

\subsection{The $2/3$ state}\label{edge:2/3}
Now we consider $\nu=2/3$ state and make a comparison with $1/3$ state.
The topological characters are shown in Fig.\ref{fig:2_3_Topological_Order_Cou}. The lower panel shows the TEE of $2/3$ state converges to  $\ln{\sqrt{3}}$, similar to the Laughlin $1/3$ state. This value of TEE shows the abelian nature of the $2/3$ state.
Meanwhile, the ES of $2/3$ state in Fig.\ref{fig:ES_Cou_b_20}(b) also exhibits chirality and degeneracy but its direction of motion is opposite to Laughlin $1/3$ state which is shown in Fig.\ref{fig:ES_Cou_b_20}(a).
This supports that the $\nu=2/3$ state is a hole-type Laughlin $1/3$ state~\cite{Girvin1984, MacDonald1990}:
\begin{align}
\nonumber
	S &= \int dxdt \frac{1}{4\pi}\partial_x\phi_1(\partial_t\phi_1-v_1\partial_x\phi_1) \\
	&-\frac{3}{4\pi}\partial_x\phi_3(\partial_t\phi_3+v_3\partial_x\phi_3)
\end{align}
where $\phi_1$ is the edge mode of $\nu=1$ component and $\phi_3$ is the edge mode of the Laughlin $\nu=1/3$ component.

In addition, the momentum polarization is shown in Fig.\ref{fig:2_3_Topological_Order_Cou}, the topological spin of $2/3$ state is $h=1/3$,
but the guiding center spin takes the opposite value $s\approx 1$ that is opposite to that of $1/3$ state, which is consistent with theoretical prediction(see Appendix\ref{apdx::shift}). From this, we also state that the topological shift is $\mathcal S=0$ for $2/3$ state, pointing to Abelian Laughlin state.

The numerically obtained chiral central charge perfectly converges to $0$ as $L_y$ increasing. The result $c\approx 0$ supports two counter-propagating $\nu=1$ and $\nu=1/3$ edge modes. This is also consistent with the exact diagonalization study on the torus geometry \cite{TwistED}.

\begin{figure}[t]
	\includegraphics[width=0.48\textwidth]{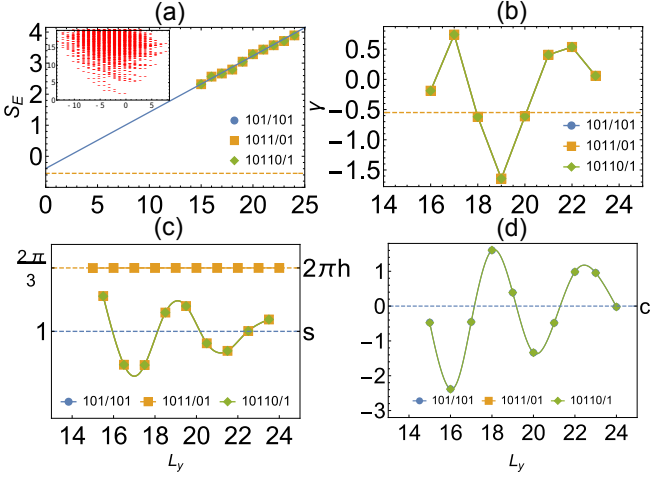}
	\caption{\textit{The topological characters of $\nu=8/3$ state for Coulomb interaction:}
		(a) The EE and (b) the extracted TEE of $\nu=8/3$ state, the yellow dashed line is $-\ln\sqrt{3}$. (c-d) The momentum polarization of $\nu=8/3$ state in different $L_y$.
		Fig.(c) are the extracted topological spin $h_a$ and guiding center spin $s$, Fig.(d) is the central charge $c$.
	}\label{fig:8_3_iDMRG}
\end{figure}
\subsection{The $2+2/3$ state}

We consider the $\nu=8/3$ state in SLL. First, the ES has shown in the inset of Fig.\ref{fig:8_3_iDMRG}(a) which shows the same ES as the ES of $2/3$ state. The typical $1,1,2,3,5,...$ counting matches the prediction of free chiral Boson.
In addition, the EE $S_E$ scales as area law in Fig.\ref{fig:8_3_iDMRG}(a), the fitting result is $S_E\approx -0.3919+0.1813L_y$. The extracted TEE is $\gamma\approx 0.3919$ which is nearly 29\% error from the theoretical prediction of $2/3$ state($\ln\sqrt{3}\approx 0.5493$), the big error comes from the finite size effect of $L_y$. 
Since the Coulomb interaction at SLL decays much slower than the lowest Landau level(LLL), resulting in a rapid increase of the bond dimension of MPO($\sim1000-2000$), so it is difficult to get convergence in large $L_y$.
On the other hand, the momentum polarization in Fig.\ref{fig:8_3_iDMRG}(c-d) tells us the $8/3$ state contains only one kind of topological spin $h_a=1/3$, consistent with the Abelian $2/3$ state. Moreover, the extracted guiding center spin $s$ and central charge $c$ are shown in Fig.\ref{fig:8_3_iDMRG}(c-d), they all oscillate and converge slowly to the theoretical value of $\nu=2/3$ state($s=1$ and $c=0$) as $L_y$ increasing\cite{note2}.
Combining the above results, we can conclude that the $8/3$ state is Abelian, which is topologically equivalent to the $2/3$ state in LLL.
\begin{figure}[b]
	\includegraphics[width=0.46\textwidth]{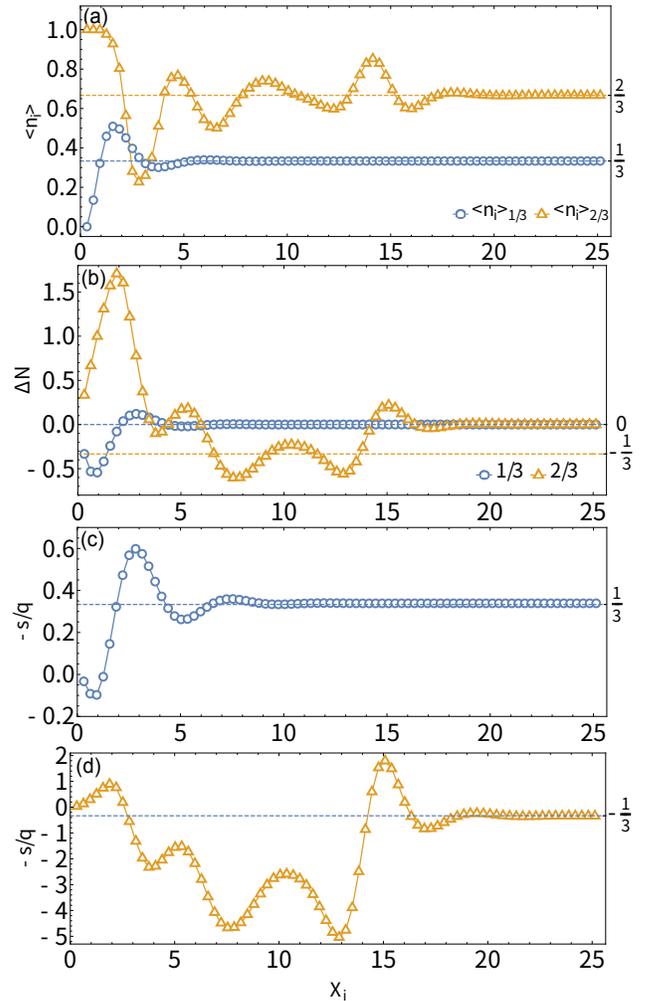}
	\caption{\textit{The egde of $\nu=1/3,2/3$ state: }
		(a) The density profile and (b) integral of Laughlin $1/3$ state(blue labels) and $2/3$ state(yellow labels). (c-d) The extracted guiding center spin from the dipole moment, where (c) for Laughlin $1/3$ state and (d) for $2/3$ state. We can see the quantity $-s/q$ conveges to $1/3$ and $-1/3$ for Laughlin $\nu=1/3$ state and $\nu=2/3$ state respectively.
		Here, we choose the Haldane pseudopotential $v_1=1.0$ and circumference of cylinder $L_y=20\ell$.
	}\label{fig:dipole_pseudo}
\end{figure}

\begin{figure*}[t]
	\includegraphics[width=0.87\textwidth]{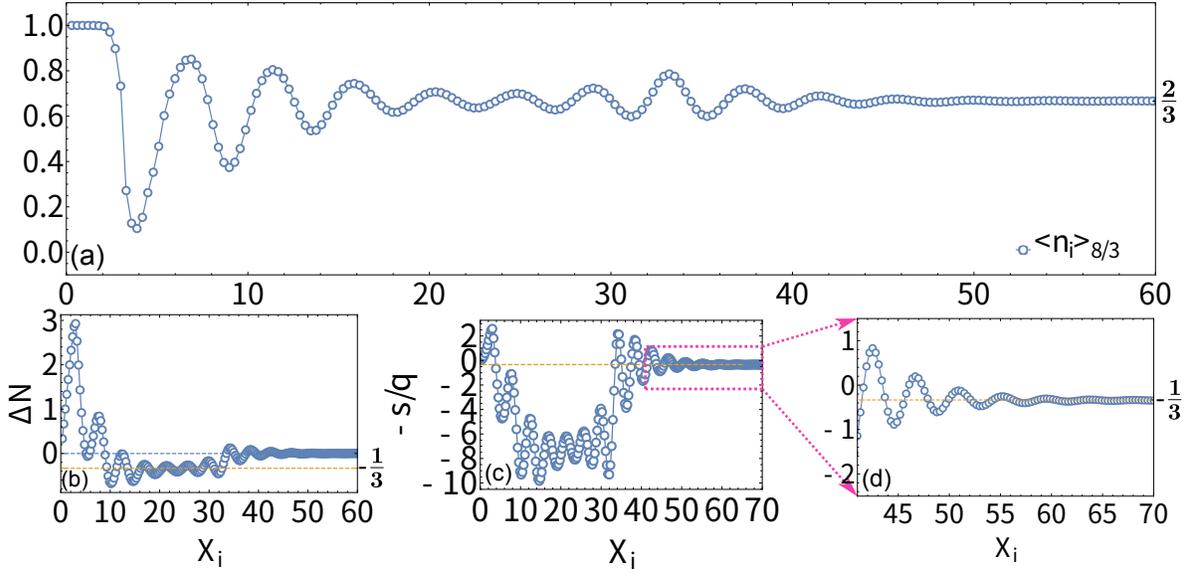}
	\caption{\textit{The "hardwall edge" of $\nu=8/3$ state:}
		(a) The density profile and (b)integral of the $\nu=8/3$ state. (c) The guiding center spin extracted from dipole moemnt. (d) A zoomed portion of (c), we can clearly see that $-s/q$ perfectly converge to $-1/3$ as the theoretical prediction.
		Here, we choose Coulomb interaction and the circumference $L_y=21$.
	}\label{fig:Hard_8_3_Interface_coulomb}
\end{figure*}

\section{Topological properties on the hardwall edge}\label{sec::edge}
Apart from the topological order in bulk, we will investigate the edge of the $1/3,2/3,8/3$ states in this section. Here, when we call the edge, it is equivalent to ``hardwall$|\nu$" interface as shown in Fig. \ref{fig::DMRGinterface} (top).

\subsection{Density profile}
The expectation of particle number operator
$\langle\hat{n}_i\rangle=\langle\hat{c}^\dagger_i\hat{c}_i\rangle$ is the most direct representation of edge. We plot the density profile of the edge (or ``hardwall" interface see Fig. \ref{fig::DMRGinterface}(top)) of $\nu=1/3$ and $\nu=2/3$ state in Fig.\ref{fig:dipole_pseudo}(a). The edge of $1/3$ state exhibits oscillating behavior, similar to Ref.\cite{RotonEdge} the density profile can be well fitted by $f_\nu(x)=C_\nu\exp{(-x/\xi_\nu)}\cos{(k_\nu x+\theta_\nu)}+\nu$, where $k_{1/3}\approx 1.546\ell^{-1}$ and $\xi_{1/3}\approx 1.573\ell$. The wave number $k_{1/3}\approx 1.546\ell^{-1}$ is close to the bulk magnetoroton minimum, agreement with the result in Ref.\cite{YoshiokaRoton1986,RotonEdge,Magnetoexciton1996}. The density profile of $2/3$ state has a $\nu=1$ integer region at the outmost. After an intermediate region, a typical oscillating behavior can be found($X_j>14\ell$) and the same fitting result as $1/3$ state($k_{2/3}\approx 1.549\ell^{-1}$ and $\xi_{2/3}\approx 1.552\ell$). The results of $8/3$ state are shown in Fig.\ref{fig:Hard_8_3_Interface_coulomb}. There is also an outmost $\nu=1$ integer region, where the typical oscillating behavior can be found in $X_j>34\ell$ and the fitting results are $k_{8/3}\approx1.517\ell^{-1}$ and $\xi_{8/3}\approx 5.482\ell$. The $2/3$ and $8/3$ edge profile is consistent with the theoretical prediction, which includes an outmost $\nu=1$ edge and an inner $\nu=1/3$ edge.

\begin{figure}[b]
	\includegraphics[width=0.46\textwidth]{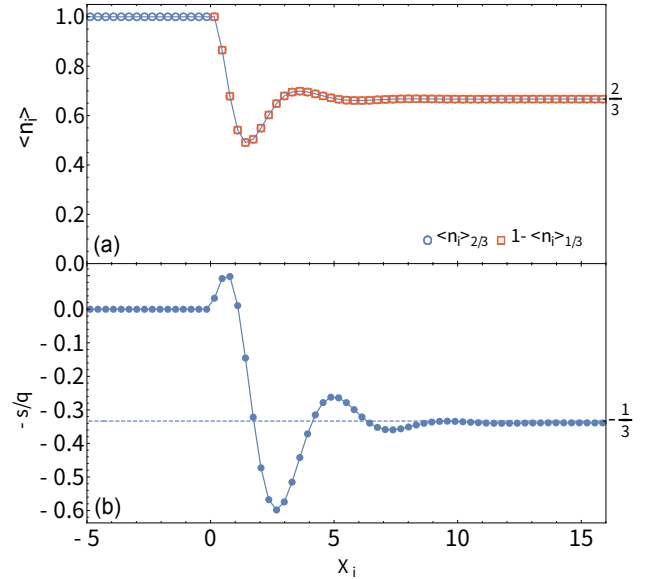}
	\caption{\textit{The density profile of $\nu=1|\nu=2/3$ interface: }
		(a) The density profile of $\nu=1|\nu=2/3$ interface(blue circle). The red square is $1-\langle\hat{n}_i\rangle_{1/3}$ where $\langle\hat{n}_i\rangle_{1/3}$ is the density of Laughlin $\nu=1/3$ state in Fig.\ref{fig:dipole_pseudo}(a). Here, we see $\nu=1|\nu=2/3$ interface is equivalent to a hole-type Laughlin $1/3$ state embedded in the integer quantum Hall $\nu=1$ background.
		(b) The extracted guiding center spin from dipole moment of $\nu=1|\nu=2/3$ interface, the results converge to $-1/3$ perfectly as $X_j$ increasing where $-s/q=-1/3$ is the theoretical prediction.
	}\label{fig:IQH1_2_3_Interface_pseudo}
\end{figure}

\begin{figure*}[t]
	\includegraphics[width=0.87\textwidth]{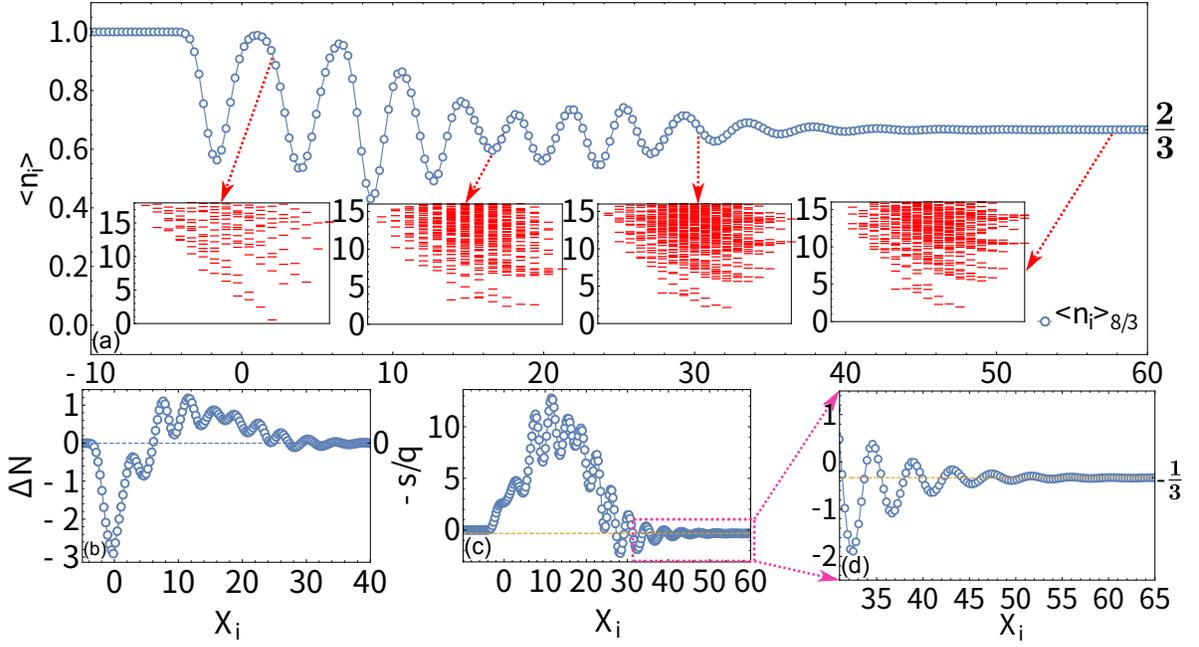}
	\caption{\textit{The $\nu=3|\nu=8/3$ interface:}
		(a) The density profile of the $\nu=3|\nu=8/3$ interaface. The left halve($X_j<0$) is $\nu=3$($\nu=1$ in the SLL) IQH state and right($X_j>0$) is $\nu=8/3$($\nu=2/3$ in the SLL). The insets show the ES in different positions, they all exhibit the same chirality. (b) The density integral of this interface. (c) The guiding center spin extracted from dipole moemnt. (d) A zoomed portion of (c), we can clearly see that $-s/q$ perfectly converge to $-1/3$ as the theoretical prediction.
		Here, we choose Coulomb interaction and the circumference $L_y=21$.
	}\label{fig:IQH_8_3_Interface_coulomb}
\end{figure*}

\subsection{Dipole moment and guiding center spin}\label{edge:dipole}
The oscilating density induces a dipole moment on the edge. Before calculating dipole moment, we should check the charge neutral conditions where density integral
$\Delta N(x) = \sum_i^x \left(\langle\hat{n}_i\rangle-\nu \right)$ should converge to $0$~\cite{YJPark2014}. This condition ensures that the dipole moment does not depend on the choice of origin. In Fig.\ref{fig:dipole_pseudo}(b) and Fig.\ref{fig:Hard_8_3_Interface_coulomb}(b), we can see $\Delta N(x)$ converges to $0$ in bulk.
The dipole moment in inhomogeneous external electric field will feel a electric force, and this force can be balanced by the viscous force. Thus the relation between Hall viscosity and dipole moment is $\eta^g_H = -\frac{\hbar}{4\pi\ell^2}\frac{s}{q} = \frac{2\pi |e| B}{L_y^2}\sum_i \left(\langle n_i\rangle-\nu \right)*i$~\cite{YJPark2014,ZhuW2020}, where we can extract the topological quantity guiding center spin $s$ via
\begin{equation}\label{dipole}
-\frac{s}{q} = \frac{8\pi^2}{L_y^2}\sum_i \left(\langle n_i\rangle-\nu \right) \times i.
\end{equation}
We have shown the results of Laughlin $1/3$ state in Fig.\ref{fig:dipole_pseudo}(c), the  quantity $-s/q$ converges to $1/3$ in the bulk, and the extracted guiding center spin is $s\approx-1$. The results of $2/3$ and $8/3$ state have shown in Fig.\ref{fig:dipole_pseudo}(d) and Fig.\ref{fig:Hard_8_3_Interface_coulomb}(c-d), the extracted guiding center spin are $s_{2/3}\approx s_{8/3}\approx1$. Both results are consistent with theoretical predictions(see Appendix\ref{apdx::shift}).

Especially, in Fig.\ref{fig:dipole_pseudo}(a-b) we can find a pair of quasi-particle($-|e|/3$) and quasi-hole($+|e|/3$) at $X_i\approx 14\ell$ and $X_i\approx 6.5\ell$ respectively. The same behavior can also be found in Fig.\ref{fig:Hard_8_3_Interface_coulomb}(b), the coordinates of quasi-particle and quasi-hole are $X_i\approx 33\ell$ and $X_i\approx 14\ell$ respectively. This pair cancels out part of the dipole moment of the outermost integer region, leaving a quantized guiding center spin.

\section{Topological properties on the interface}\label{sec::interface}

The interface between different quantum Hall states is a useful geometry to study the edge physics, which can be realized in experiments.
Especially, recent measurements on the interface between $\nu=1$ and $\nu=2/3$ state demonstrate the  counterpropagating chiral channels and upstream neutral edge mode  ~\cite{NovelDutta2021,Cohen2019,Grivnin2014}.
Motivated by these experimental progresses, in this section, we study the interface between integer quantum Hall states and FQH states. The numerical details of constructing interface in DMRG is the ``cut-and-glue" scheme, which has been shown in Sec.\ref{cons_interface}.

\begin{figure*}[t]
	\includegraphics[width=0.87\textwidth]{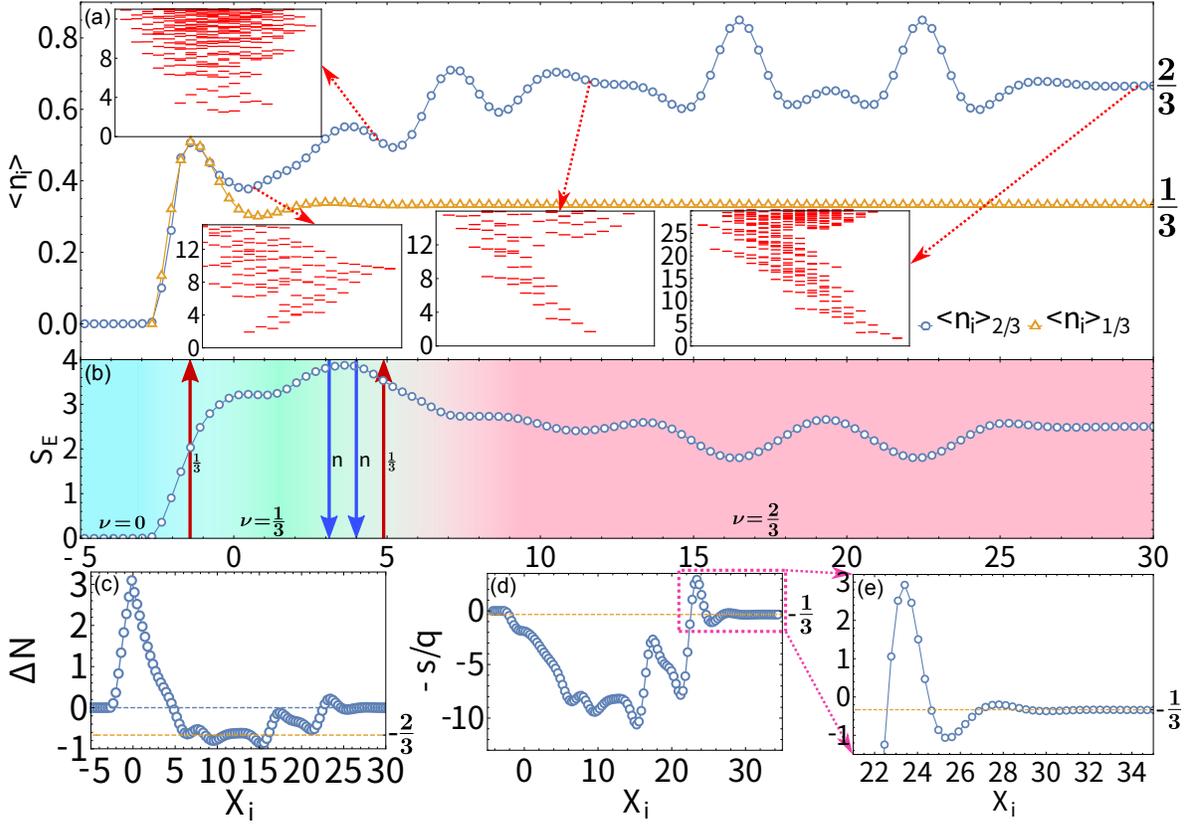}
	\caption{\textit{The $\nu=0|\nu=2/3$ interface:}
	(a) The density profile of the $\nu=0|\nu=2/3$ interaface. The left halve($X_j<0$) is $\nu=0$ vacuum state and right($X_j>0$) is $\nu=2/3$ state. The insets show the ES in different positions. The ES in region $X_i\sim 0\ell$ and $X_i>10\ell$ have opposite chirality, which means a small region of Laughlin $1/3$ has emerged. The ES in upper left conner($X_i\approx 4.39\ell$) have particle-hole symmetry, which is the evidence of the existence of the interface between $1/3$ state and its particle-hole conjugation($2/3$ state). (b) The EE of this interface, and the background is the schematic diagram of this interface, the red and blue arrows represent the $1/3$ chiral Boson and neutral mode respectively. (c) The density integral of this interface. (d) The guiding center spin extracted from dipole moment. (e) A zoomed portion of (d), we can clearly see that $-s/q$ perfectly converge to $-1/3$ as the theoretical prediction.
	Here, we choose Haldane pseudopotential $v_1=1.0$ and the circumference $L_y=20$.
	}\label{fig:Vacuum_2_3_Interface_pseudo}
\end{figure*}

\subsection{The $\nu=1|\nu=2/3$ interface}\label{1|2/3_interfaec}

Let us recall that, in the Sec. \ref{edge:2/3}, the $\nu=2/3$ state supports two counterpropagating chiral channels: a downstream $\nu=1$ electron channel, and an upstream $\nu=1/3$ channel.
On the interface between $\nu=1$ and $\nu=2/3$ it is expected to have only one $1/3$ edge mode, since the integer edge mode of $2/3$ state is gapped out by $\nu=1$ state. The single $1/3$ edge mode is expected to exhibit a $1/3$ two-terminal conductance, which has been seen in experiment~\cite{NovelDutta2021}. Similarly, a $\nu=3|\nu=8/3$ interface, the two integer edge modes in lower Landau level gap out each other, equivalent to a $\nu=1|\nu=2/3$ interface.

The blue line in Fig.\ref{fig:IQH1_2_3_Interface_pseudo}(a) is the density $\langle\hat{n}_i\rangle_{2/3}$ of this interface,
and the left half($X_j<0$) is $\nu=1$ state and right($X_j>0$) is $\nu=2/3$ state. We have found two interesting points from the density profile. First, there is no charge ``leaking" from $\nu=1$ to $\nu=2/3$, reflecting the incompressible natural of $\nu=2/3$ state. Second, we compare the density profile with the hole-Laughlin state $ 1-\langle\hat{n}_i\rangle_{1/3}$, where $\langle\hat{n}_i\rangle_{1/3}$ is the density of Laughlin $1/3$ state as shown in Fig.\ref{fig:dipole_pseudo}(a). It is found that the density profile around $\nu=1|\nu=1/3$ interface perfectly matches the hole-Laughlin state. It strongly indicates the $\nu=1|\nu=2/3$ interface  forms a hole-type Laughlin $\nu=1/3$ state.

Similar to Sec.\ref{edge:dipole}, the balance condition on the interface requires to form a quantized electric dipole moment ~\cite{ZhuW2020}
\begin{equation}\label{interface:dipole}
-\left(\frac{s^R}{q^R}-\frac{s^L}{q^L}\right) = \frac{8\pi^2}{L_y^2}\sum_i \left(\langle n_i\rangle-\nu_i \right) \times i.
\end{equation}
where $\nu_i=\theta(i)\times 1+\theta(-i)\times 2/3$ is the filling of each half and $\theta(x)$ is step function. In the left region $s^L=0$ for $\nu=1$ state, and in the right region $s^R/q^R=1/3$ for $\nu=2/3$ state (as shown in Sec. \ref{edge:2/3}), we expect  $-1/3$ in Eq.\ref{interface:dipole} on the $\nu=1|\nu=2/3$ interface.
In Fig.\ref{fig:IQH1_2_3_Interface_pseudo}(b), we have shown the numerical result of dipole moment where the results converge to $-1/3$ perfectly as $X_j$ increasing.

Now we consider the similar $\nu=3|\nu=3/8$ interface or $\nu=1|\nu=2/3$ interface in SLL, where the results have shown in Fig.\ref{fig:IQH_8_3_Interface_coulomb}. In Fig.\ref{fig:IQH_8_3_Interface_coulomb}(a), we have shown the density of $\nu=3|\nu=2/3$ interface, the insets are ES at different positions. In Fig.\ref{fig:IQH_8_3_Interface_coulomb}(b), the density integral converges to $0$ indicates the neutral condition. Using Eq.(\ref{interface:dipole}) and choosing $\nu_{i<0}=1, \nu_{i>0}=2/3$, we have shown the results in Fig.\ref{fig:IQH_8_3_Interface_coulomb}(c-d). The result converges to $-1/3$ perfectly as $X_j$ increasing, which is consistent with the result of $\nu=1|\nu=2/3$ interface in LLL. This result shows that the $\nu=8/3$ state has the same guiding center spin as $\nu=2/3$ state. It can be seen as an evidence of that $\nu=8/3$ and $\nu=2/3$ states have the same topological order.

\begin{figure*}
	\includegraphics[width=0.87\textwidth]{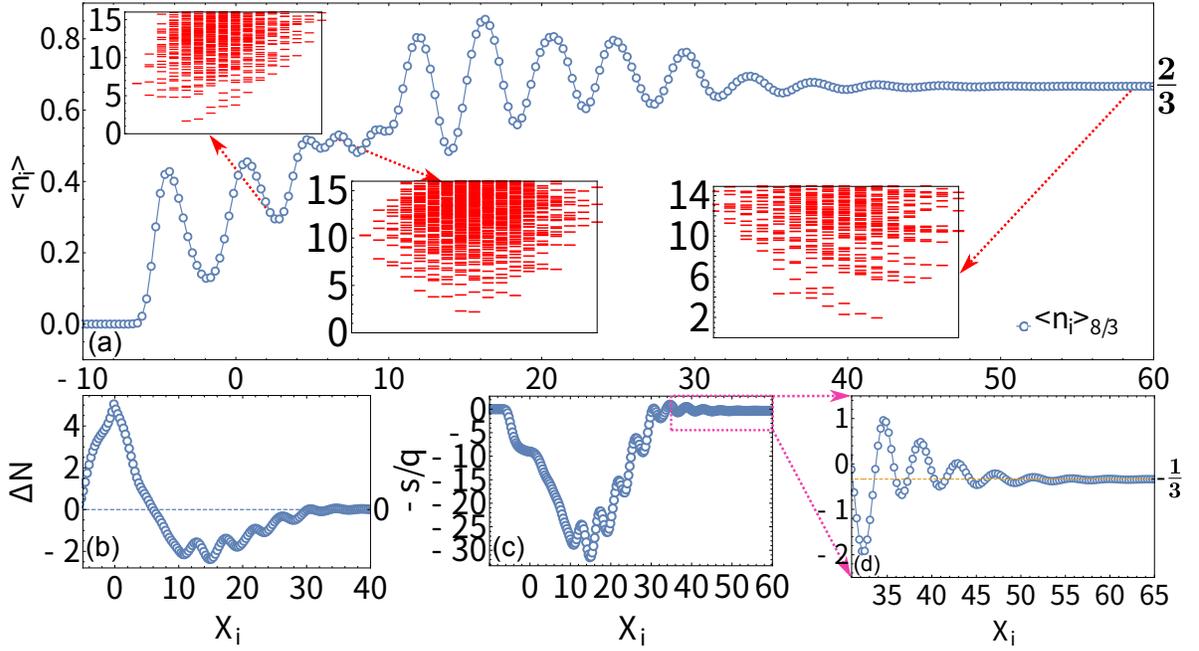}
	\caption{\textit{The $\nu=2|\nu=8/3$ interface: }
		(a) The density profile of the $\nu=2|\nu=8/3$ interaface. The left halve($X_j<0$) is $\nu=2$($\nu=0$ in the SLL) state and right($X_j>0$) is $\nu=8/3$($\nu=2/3$ in the SLL) state. The insets show the ES in different positions. (b) The density integral of this interface. (c) The guiding center spin extracted from dipole moment. (d) A zoomed portion of (c), we can clearly see that $-s/q$ perfectly converge to $-1/3$ as the theoretical prediction.
		Here, we choose Coulomb interaction and the circumference $L_y=21$.
	}\label{fig:Vacuum_8_3_Interface_coulomb}
\end{figure*}

\subsection{The $\nu=0|\nu=2/3$ interface }\label{edgeReconstru}
We concentrate now on the interface between $\nu=0$ and $\nu=2/3$ which is equivalent to the  $\nu=2|\nu=8/3$ interface. We stress that the difference between the $\nu=0|\nu=2/3$ interface and the hardwall edge of $\nu=2/3$ (see Sec.\ref{edge:2/3}): the edge can be seen as a special interface with a  infinite confining potential, under this potential the electrons never cross the interface to $\nu=0$ region, as a result an outmost $\nu=1$ region is formed (see Fig.\ref{fig:dipole_pseudo}(a)). Here, we choose the confining potential to be $0$, therefore electrons cross the interface to $\nu=0$ region and the original $\nu=1$ region collapses, thus the edge reconstruction occurs. The former study~\cite{Meir1994} suggested that a $\nu=1/3$ stripe will appear on the outmost. As a result, two new counterpropogating $\nu=1/3$ edge mode could be added to the outside of the original (in order from edge to bulk, $1/3, -1/3, 1, -1/3$)~\cite{Meir2013,NP2/3_2017}. The former study proposed that the inner three edge modes should renormalize to a single $1/3$ charge mode and two opposite neutral modes.

In Fig.\ref{fig:Vacuum_2_3_Interface_pseudo}(a), the blue line is the density profile of $\nu=0|\nu=2/3$ interface, where the left half($X_i<0$) is vacuum $\nu=0$ and right($X_i>0$) is $\nu=2/3$. We identify the density in $X_i<0$ region forms a typical Laughlin $\nu=1/3$ edge, by comparing the density profile with that of Laughlin $1/3$ state (the yellow line) (same data from Fig.\ref{fig:dipole_pseudo}(a)). The inset at the bottom left of Fig.\ref{fig:Vacuum_2_3_Interface_pseudo}(a), is the entanglement spectrum at $X_i\approx0.47\ell$, which shows the same chirality and countings as Laughlin $\nu=1/3$ state. Combining these two evidences, we can conclude that there is a $1/3$ edge mode at the outmost region($-3\ell<X_i<0$). Moving to $X_i\approx4.39\ell$, the entanglement spectrum shown in the upper left conner of Fig.\ref{fig:Vacuum_2_3_Interface_pseudo}(a), it exhibits the particle-hole symmetry, signaling a crossover regime. Moreover, the entanglement spectrum on the right side of this point all show the same chirality as $\nu=2/3$ state.
Based on these observations, we speculate that, around $X_i\approx4.39\ell$ and $\langle\hat{n}_i\rangle\approx0.5$ it emerges a  $\nu=1/3|\nu=2/3$ interface.
This is partially consistent with the prediction in Ref. \cite{Gefen2013} and experimental observation \cite{NP2/3_2017}.

Now, we focus on the dipole moment of the $\nu=0|\nu=2/3$ interface. The density integral
$\Delta N(x) = \sum_i^x \left(\langle\hat{n}_i\rangle-\nu \right)$ is shown in Fig.\ref{fig:Vacuum_2_3_Interface_pseudo}(c) and we obtain $\Delta N(X_i)=0$ for large $X_i$. We further calculate the dipole moment of this interface by Eq.(\ref{interface:dipole}).
In Fig.\ref{fig:Vacuum_2_3_Interface_pseudo}(d) and (e), we have shown the numerical result of Eq.\ref{interface:dipole} where $-s^{\nu=2/3}/q$ converge to $-0.3387$ as $X_j$ increasing, and extracted $s^{\nu=2/3}\approx1.0162$. Thus, the dipole moment on the $\nu=0|\nu=2/3$ interface is same as $\nu=2/3$ edge, because both $\nu=0$ and vaccum have the same guiding center spin $s^{\nu=1}=0$.

Finally, we consider the $\nu=2|\nu=8/3$ interface, where the integer part gap out each other, leaving the $\nu=0|\nu=2/3$ in the SLL. Comparing Fig.\ref{fig:Vacuum_2_3_Interface_pseudo} and Fig.\ref{fig:Vacuum_8_3_Interface_coulomb}, we can find that the $\nu=2|\nu=8/3$ interface is similar to the $\nu=0|\nu=2/3$ interface, except for the different penetration depths(depend on the nonuniversal correlation length). One can find there is also a small region of $2+1/3$ Laughlin state between $\nu=2$ and $\nu=8/3$, i.e. the edge reconstruction happens.

\begin{figure*}
\includegraphics[width=0.87\textwidth]{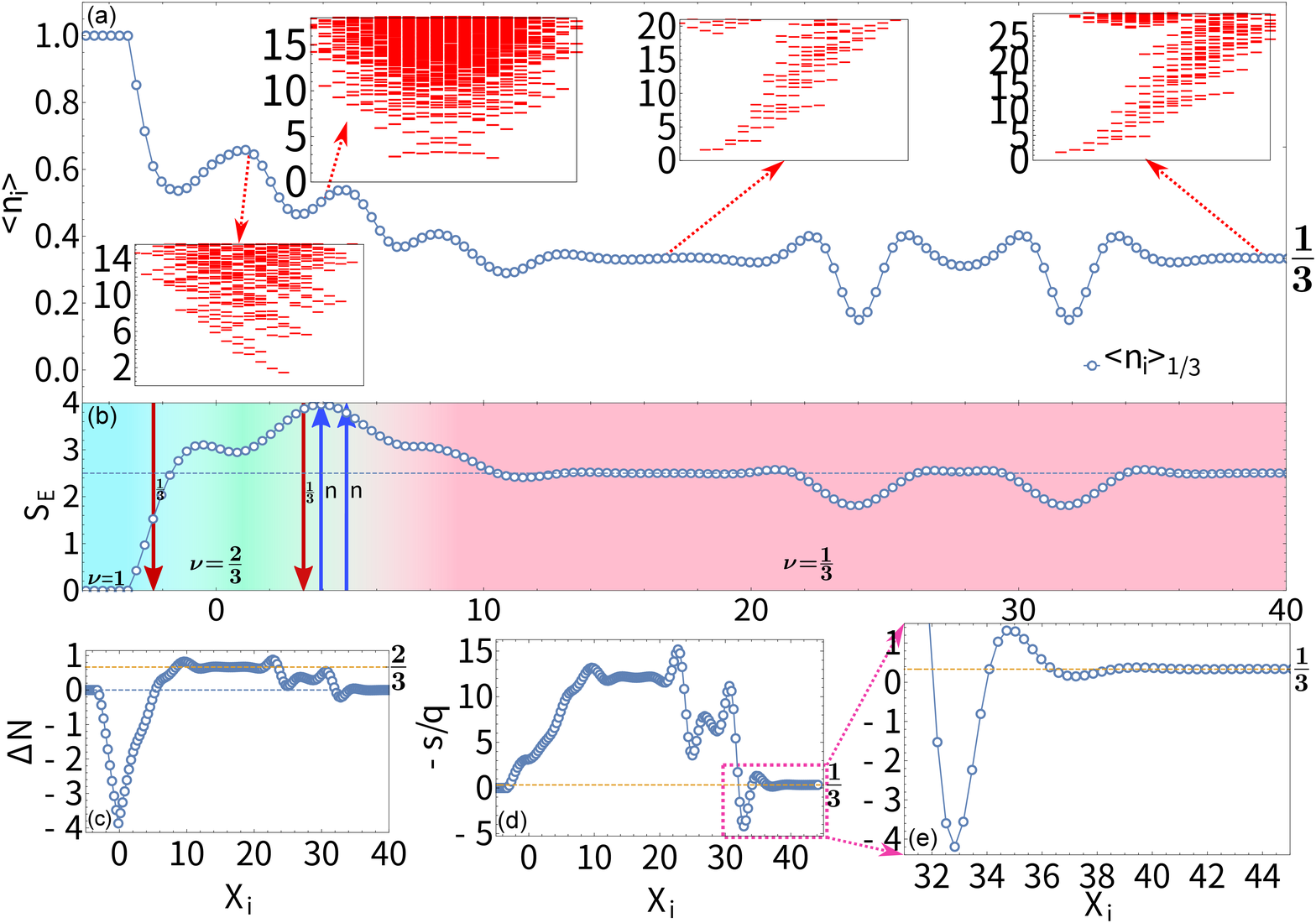}
\caption{\textit{The $\nu=1|\nu=1/3$ interface: }
	(a) The density profile of the $\nu=1|\nu=1/3$ interaface. The left halve($X_j<0$) is $\nu=1$ state and right($X_j>0$) is $\nu=1/3$ state. The insets show the ES in different positions. The ES in region $X_i~0\ell$ and $X_i>10\ell$ have opposite chirality, which means a small region of $2/3$ has emerged. The ES in upper left conner($X_i\approx 4\ell$) have particle-hole symmetry, which is the evidence of the existence of the interface between $1/3$ and $2/3$ state. (b) The EE of this interface, and the background is the schematic diagram of this interface, the red and blue arrows represent the $1/3$ chiral Boson and neutral mode respectively. (c) The density integral of this interface. (d) The guiding center spin extracted from dipole moemnt. (e) A zoomed portion of (d), we can clearly see that $-s/q$ perfectly converge to $1/3$ as the theoretical prediction.
	Here, we choose Haldane pseudopotential $v_1=1.0$ and the circumference $L_y=20$.
}\label{fig:IQH_1_3_Interface_pseudo}
\end{figure*}

\subsection{The $\nu=1|\nu=1/3$ interface}\label{1|1/3_interfaec}
At the end of this section, we consider the $\nu=1|\nu=1/3$ interface which is also studied in Ref.\cite{IQH_Edge_re,Lai2013}. Since the $\nu=1|\nu=1/3$ interface is the particle-hole conjugation of the $\nu=0|\nu=2/3$ interface. Meanwhile, in Sec.\ref{edgeReconstru}, we find the edge reconstruction in the $\nu=0|\nu=2/3$ interface, resulting in a small $\nu=1/3$ region between $\nu=0$ an $\nu=2/3$, thus we look forward to discovering a small $\nu=2/3$ region on  the $\nu=1|\nu=1/3$ interface. Fortunately, in the insets of Fig.\ref{fig:IQH_1_3_Interface_pseudo}(a), we find the ES around $X_i\approx1.5\ell$ exhibits the achiral property, indicates a small $\nu=2/3$ region emerged. Following the discussion in Sec.\ref{edgeReconstru}, the outmost edge mode comes from the $\nu=1|\nu=2/3$ interface, i.e a single $1/3$ charge mode(Fig.\ref{fig:IQH_1_3_Interface_pseudo}(b)). As for the inside, a $1/3$ charge mode and two opposite neutral modes(Fig.\ref{fig:IQH_1_3_Interface_pseudo}(b)) come from the $\nu=2/3|\nu=1/3$ interface. In Fig.\ref{fig:IQH_1_3_Interface_pseudo}(c-d), the extracted guiding center spin is $\Delta s= s^{\nu=1/3}-s^{\nu=1}=-1$. Once again, it supports that the dipole moment is a quantized quantity on the interface, protected by intrinsic topology.

\section{Conclusion}
In this paper, we investigate the topological properties of $\nu=2/3$ and $8/3$ state, including the bulk and edge physics.
On one hand, via the entanglement spectra and momentum polarization calculations, we identify both of $2/3,8/3$ states are Abelian type, which share the same topology as hole-conjugated Laughlin $1/3$ state.
On the other hand, we illustrate the interface made of $\nu=2/3 (8/3)$ state and $\nu=0(2)$. 
Crucially, our method is able to demonstrate the edge reconstruction directly, i.e. it identifies the interface $\nu=0|\nu=2/3$ contains a $1/3$ edge mode and a group of edge modes inherited from $\nu=1/3|\nu=2/3$ interface, which are spatially separated located on the interface.
Throughout this paper, we find that the dipole moment living on the interface is a quantized quantity (equivalent to guiding-center spin or topological shift), which is protected by non-trivial topology, even though charge density is non-universal depending on the interaction details.
Lastly, in the case of hardwall edge (sharp confining potential), the edge reconstruction doesnot happen, and the edge state forms a group with outer $\nu=1$ integer quantum Hall mode and an inner counter-propagating $1/3$ mode.

These results open a door to looking at fractional states and the behavior of edge modes, especially in cases where reconstruction at the edge takes place, leading
to formation of counter-propagating mode. For example, it was experimentally identified \cite{Bid2009,NP2/3_2017} and then theoretically proposed \cite{Gefen2013} that,
reconstructed edge at $2/3$ state consists of an outermost $1/3$ channel and the inner composite channels including neutral modes. Our current results indeed support the above picture (on the interface or smooth edge potentials), which has not been achieved by numerical simulations before. Thus, we believe our work paves the road to a more completed understanding of the fractional quantum Hall states, and it can be extended to more fillings without any barrier.

\section{ACKNOWLEDGMENTS}
WZ thanks F. D. M. Haldane for educating the emergent dipole moment around the FQH edge. We also thank D. N. Sheng for collaborating on a related project.
LDH and WZ are supported by project 11974288 from NSFC, the Key R\&D Program of Zhejiang Province (2021C01002)  and the foundation from Westlake University.
We thank Westlake University HPC Center for computation support.

\begin{appendices}

	\section{The details of many-body Hamiltonian}\label{apdx:Hamiltonian}
	In this appendix, we will show more details about the derivation of Hamiltonian and the Fourier transformation of the modified Coulomb
	interaction.
	\subsection{Hamiltonian}
	First, the Landau level wavefunction is
	\begin{equation}\label{apdx:LL-cylinder}
	\psi_{n,j}(x,y) = \frac{\mathrm{e}^{-i \frac{X_j}{\ell^2} y-\frac{(x-X_j)^2}{2\ell^2}}}{\sqrt{2^n n! \sqrt{\pi} \ell L_y}}
	H_n\left( \frac{x-X_j}{\ell}\right)  .
	\end{equation}
	If we project the electron-electron interaction into single Landau level, the second quantization form of Hamiltonian can be expressed
	by
	\begin{equation}
	H_I = \sum_{j_1,j_2,j_3,j_4}^{N_\phi} A^n_{j_1,j_2,j_3,j_4} a^\dagger_{n,j_1}a^\dagger_{n,j_2}a_{n,j_3}a_{n,j_4}
	\end{equation}
	where the interaction matrix element:
	\begin{align}\label{apdx:A}\nonumber
	A^n_{j_1,j_2,j_3,j_4} =& \frac12 \int \mathrm{d}\bm{r}_1 \int \mathrm{d}\bm{r}_2
	\psi_{N,j_1}^*(\bm{r}_1)\psi_{N,j_2}^*(\bm{r}_2) \\
	&V(\bm{r}_1,\bm{r}_2) \psi_{N,j_3}(\bm{r}_2)\psi_{N,j_4}(\bm{r}_1).
	\end{align}
	The two-body interaction in real space satisfy $V(\bm{r}+b\hat{\bm{e}}_y) = V(\bm{r})$, we can
	rewrite as:
	\begin{equation}\label{apdx:Vr}
	V(\bm{r}) = \frac1L_y \int_{-\infty}^{\infty}\mathrm{d}q_x\sum_{q_y} V(q)
	\mathrm{e}^{i\bm{q}\cdot\bm{r}}
	\end{equation}
	where $q_y=\frac{2\pi t}{L_y}, t\in\mathbb{Z}$ is the momentum along $y$-direction  and
	\begin{align}\label{apdx:Vq}
	V(q) = \int_{-\infty}^{\infty}\mathrm{d}x\int_0^{L_y} \mathrm{d}y V(\bm{r})
	\mathrm{e}^{i\bm{q}\cdot\bm{r}}
	\end{align}
	Now we can caculate $A^N_{j_1,j_2,j_3,j_4}$ by substituting Eq.(\ref{apdx:Vr}) into Eq.(\ref{apdx:A}):
	\begin{align}
	A^N_{j_1,j_2,j_3,j_4} = \frac{1}{2L_y}\int_{-\infty}^{\infty}\mathrm{d}q_x\sum_{q_y}V(q)
	I^N_{j_1,j_4}(\bm{q})  I^N_{j_2,j_3}(-\bm{q})
	\end{align}
	where $I^N_{s,s\prime}(\bm{q}) = \int\mathrm{d}\bm{r}\psi_{N,s}^*\psi_{N,s\prime}\mathrm{e}^{i\bm{q}\cdot\bm{r}}$ is a useful integral:
	\begin{align}\nonumber
	I^n_{s,s\prime}(\bm{q}) = &
	\frac{1}{2^n n! \sqrt{\pi} \ell}\int_{-\infty}^{\infty}\mathrm{d}x
	\mathrm{e}^{-\frac{1}{\ell^2} \left[
		x-\frac12 (2X_s-q_y\ell^2+iq_x\ell^2)		
		\right]^2}
	\\\nonumber
	& \times H_n((x-X_s)/\ell) H_n((x-X_{s\prime})/\ell)
	\\\nonumber
	& \times
	\mathrm{e}^{
		\frac{1}{4\ell^2} (2X_s-q_y\ell^2+iq_x\ell^2)^2
		-\frac{X_s^2}{2\ell^2}-\frac{(X_s-q_y\ell^2)^2}{2\ell^2}} \delta_{t,s-s\prime} \\
	= &L_n\left(\frac12 q^2\ell^2\right)
	\mathrm{e}^{-\frac14 q^2\ell^2+iq_xX_s-\frac{iq_xq_y\ell^2}{2}} \delta_{t,s-s\prime}
	\end{align}
	We first integrate the variable $y$, and then use a special function integral relation $\int_{-\infty}^{\infty}\mathrm{d}x\mathrm{e}^{-x^2}H_m(x+y)H_n(x+z)=
	2^n\sqrt{\pi}m!z^{n-m}L_m^{n-m}(-2yz)~~m\leq n$.
	
	Finally, we have:
	\begin{align}\nonumber
	A^n_{j_1,j_2,j_3,j_4} = & \frac{1}{2b}\int_{-\infty}^{\infty}\mathrm{d}q_x\sum_{q_y}V(q)
	I^n_{j_1,j_4}(\bm{q}) I^n_{j_2,j_3}(-\bm{q}) \\\nonumber
	= &  \frac{1}{2L_y}\int_{-\infty}^{\infty}\mathrm{d}q_x\sum_{q_y}V(q)
	\left[ L_N\left(\frac12 q^2\ell^2\right) \right]^2
	\\
	\times &\mathrm{e}^{-\frac12 q^2\ell^2+iq_x(j_1-j_3)\frac{2\pi\ell^2}{L_y}} \delta_{q_y,\frac{2\pi (j_1-j_4)}{L_y}}\delta_{j_1+j_2,j_3+j_4}
	\end{align}

	\subsection{Fourier transformation of modified Coulomb interaction}
	The modified Coulomb interaction in real space is $V(\bm{r}) = \frac{e^2}{\epsilon}\frac{\mathrm{e}^{-r^2/\xi^2}}{r}$, since the periodicity along $y$-direction, is can be written as:
	
	\begin{equation}
	V(\bm{r}) = \frac{e^2}{\epsilon}\sum_{t\in\mathbb{Z}} \frac{\mathrm{e}^{-r^2/\xi^2}}{|\bm{r}+tb\hat{\bm{e}}_y|}
	=\frac1b \int_{-\infty}^{\infty}\mathrm{d}q_x\sum_{q_y} V(\bm{q})
	\end{equation}
	where $\bm{r}\cdot\hat{\bm{e}}_y\in[0,L_y)$ and $q_y=\frac{2\pi t}{L_y}, t\in\mathbb{Z}$. We have
	\begin{align}\nonumber
	V(\bm{q}) &=  \frac{e^2}{\epsilon}\int_{-\infty}^{\infty}\mathrm{d}x\int_{-\infty}^{\infty}\mathrm{d}y
	\frac{\mathrm{e}^{-r^2/\xi^2}}{r}\mathrm{e}^{-i\bm{q}\cdot\bm{r}} \\\nonumber
	&=  \frac{e^2}{\epsilon}\int_{0}^{\infty}\mathrm{d}r\int_{0}^{2\pi}\mathrm{d}\theta
	\frac{\mathrm{e}^{-r^2/\xi^2}}{r}\mathrm{e}^{-iqr\cos{\theta}} \\\nonumber
	&= \frac{e^2}{\epsilon}\int_{0}^{\infty}\mathrm{d}r \mathrm{e}^{-r^2/(q^2\xi^2)}J_0(r) \\
	&= \frac{e^2}{\epsilon}\frac{\xi}{2} \sqrt{\pi} \exp{\left(-\frac{q^2\xi^2}{8}\right)}
	I_0\left(\frac{q^2\xi^2}{8}\right)
	\end{align}
	where $\int_0^\infty\mathrm{d}r \mathrm{e}^{-r^2/\xi^2}J_0(r) = \frac{\xi}{2}\sqrt{\pi}
	\exp{(-\xi^2/8)}I_0(\xi^2/8)$ and $I_0(z) = \frac{1}{\pi}\int_0^{\pi}\exp{(z\cos{\theta})}\cos{(n\theta)}\mathrm{d}\theta$ is the \textit{1st modified Bessel function} $I_n(z)=\frac{1}{\pi}\int_0^\pi \mathrm{d}\theta
	\exp{(z\cos{\theta})}\cos{(n\theta)} $. If we choose $\xi\rightarrow\infty$, the modified Coulomb interaction with regulated length $\xi$ will come back the original Coulomb interaction
	\begin{equation}
	\lim\limits_{\xi\rightarrow\infty}V(\bm{q}) = \frac{e^2}{\epsilon} \frac{1}{q}
	\end{equation}
	where we have used the asymptotic approximation of the first kind of modified Bessel function(for large values of $x$)
	\begin{equation}
	I_n(x) = \frac{\mathrm{e}^x}{\sqrt{2\pi x}}\left[ 1-\frac{4n^2-1}{8x}+O\left( \frac{1}{x^2}\right)  \right] .
	\end{equation}

    \subsection{ Pseudopotential Hamiltonian }
In general, the $V(q)$ in Eq.(\ref{apdx:Vr}) can be expanded with Laguerre polynomials
\begin{equation}
	V(\bm q)=\sum_{l} v_lL_l(q^2)
\end{equation}
where $L_l(x)$ is Laguerre polynomial and $v_l$ are some tunable parameters.

Here, we will show that this potential is a short-range interaction, using Eq.(\ref{apdx:Vr})
\begin{align}\nonumber
	V(\bm{r}) &= \frac1L_y \int_{-\infty}^{\infty}\mathrm{d}q_x\sum_{q_y} \sum_{l} v_lL_l(q^2)
\mathrm{e}^{i\bm{q}\cdot\bm{r}}\\\nonumber
&=  \sum_{l} v_lL_l(-\bm{\nabla}_{\bm r}^2)
\int_{-\infty}^{\infty}\mathrm{d}q_x\frac1L_y\sum_{q_y} \mathrm{e}^{i\bm{q}\cdot\bm{r}}\\
&=  \sum_{l} v_lL_l(-\bm{\nabla}_{\bm r}^2)\delta^2(\bm r).
\end{align}
This is the model Hamiltonian of Laughlin $\nu=1/q$ wavefunction $\Psi_q(\{\bm r_i\})$
\begin{equation}
\int\prod_i d^2\bm r_i \Psi_q^*(\{\bm r_i\})\sum_{j,k}V(\bm r_j-\bm r_k)\Psi_q(\{\bm r_i\})=0
\end{equation}
if we choose
\begin{equation}
\left\{
\begin{aligned}
v_l&= 1 \quad (l<q)\\
v_l&=0 \quad (l\ge q)
\end{aligned}
\right.
\end{equation}
Since Laguerre polynomials are complete, by selecting the appropriate $v_l$'s, we can construct some special potential $V(\bm r)$, and $v_l$ is called Haldane pseudopotential in order $l$.

\begin{figure}[t]
	\includegraphics[width=0.46\textwidth]{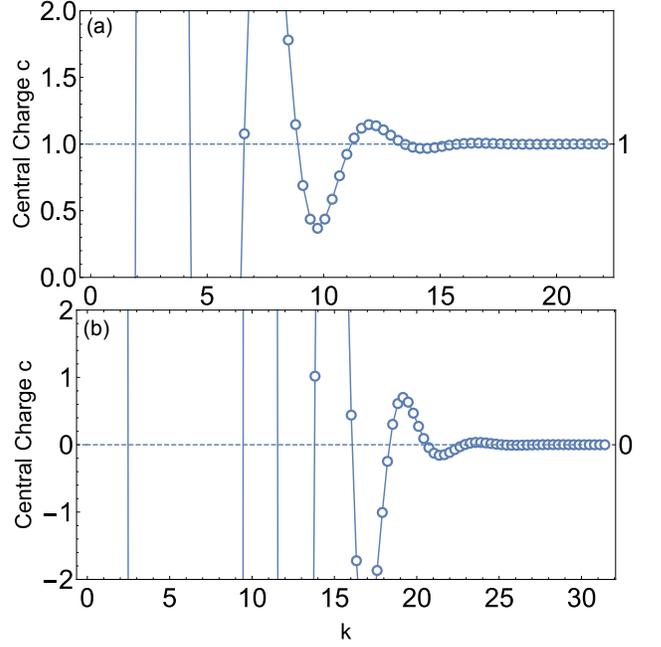}
	\caption{\textit{The central charge $c$ of $\nu=1/3,2/3$ state: }
		The extracted central charge $c$ of (a) Laughlin $\nu=1/3$ state and (b) $\nu=2/3$ state from density profile.
		The results are calculated by Eq.(\ref{apdxEQ::cc}), the density profile are the same data as in Fig.\ref{fig:dipole_pseudo} and the guiding center spin in Eq.(\ref{apdxEQ::cc}) are chosen to be $s_{1/3}=-1,s_{2/3}=1$. The extracted central charge of $\nu=1/3,2/3$ states are $c=0.999931,0.000066$ respectively.
	}\label{Afig:centralcharge}
\end{figure}
\section{The theoretical prediction of topological shift}\label{apdx::shift}
The topological shift was first introduced by Wen and Zee\cite{Zee1992}, which is a shift in the relation between $N_e$ and $N_\phi$
\begin{equation}
	N_\phi=\nu^{-1}N_e-\mathcal{S}(1-g)
\end{equation}
where $g$ is the genus of manifold. We can easily find the relations $N_\phi=\nu^{-1}N_e-\mathcal{S}$ for sphere geometry and $N_\phi=\nu^{-1}N_e$ for torus geometry. In Ref.\cite{Zee1992}, the topological shift can be determined by $K$ matrix
\begin{equation}
	\mathcal{S}=\frac{2}{\nu}\bm{q}^TK^{-1}\bm{s}
\end{equation}
where $\bm{q}$ and $\bm s$ are charge and spin vector. For Laughlin $\nu=1/q$ state in LLL, where $K=q, \bm q=1$ and $\bm s = q/2$, we have $\mathcal{S}=q$. As for the hole-type Laughlin state in filling $\nu=1-1/q$ in LLL, $K, \bm q, \bm s$ are
\begin{equation}
K=\left(
\begin{array}{cc}
1&1\\1&1-q
\end{array}\right) \quad
\bm q=\left(
\begin{array}{c}
1\\0
\end{array}\right) \quad
\bm s=\frac 12\left(
\begin{array}{c}
1\\1-q
\end{array}\right)
\end{equation}
thus we have $\mathcal{S}=0$. Using the relations in Eq.(\ref{Eq:viscosity}), we have $s^{\frac 1q}=\frac{1-q}{2}$ and $s^{1-\frac{1}{q}}=\frac{q-1}{2}=-s^{\frac{1}{q}}$. Now we consider the results in higher Landau level, the $K$ matrix and charge vector $\bm q$ donot change, but the components of spin vector $s_I$ should change as the Landau level index $n(n=0,1,2,\cdots)$
\begin{equation}
	s_I \rightarrow s_I+n
\end{equation}
now we have the final results
\begin{equation}
\begin{aligned}
&\mathcal{S}^{n+\frac{1}{q}}=q+2n  &s^{n+\frac 1q}=\frac{1-q}{2}\\
&\mathcal{S}^{n+1-\frac{1}{q}}=2n  &s^{n+1-\frac{1}{q}}=\frac{q-1}{2}
\end{aligned}
\end{equation}
We can clearly see that the guiding center spin of Laughlin $\nu=n+\frac{1}{q}$ state and its particle-hole conjugated $\nu=n+1-\frac{1}{q}$ state take opposite values.
 When we choose $q=3$, we have the theoretical predictions of Laughlin $\nu=n+\frac{1}{3}$ and $\nu=n+\frac{2}{3}$ state
 \begin{equation}
 \begin{aligned}
 &\mathcal{S}^{n+\frac 13}=3+2n  &s^{n+\frac 13}=-1\\
 &\mathcal{S}^{n+\frac 23}=2n  &s^{n+\frac 23}=1
 \end{aligned}
 \end{equation}
 
 \begin{figure}[t]
 	\includegraphics[width=0.46\textwidth]{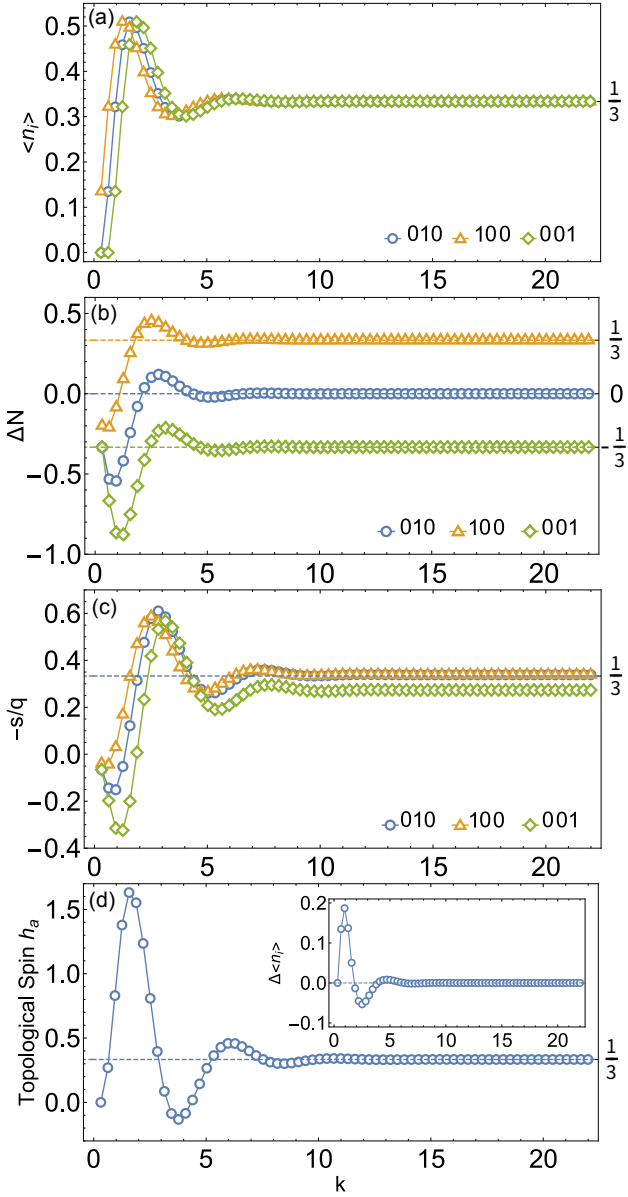}
 	\caption{\textit{The ``hardwall edge" of different topological sectors of Laughlin $\nu=1/3$ state: } (a) The density profile $\langle\hat{n}_i \rangle$, (b) density integral $\Delta N$ and (c) extracted $-s/q$ of different topological sectors of Laughlin $\nu=1/3$ state. (d) The topological spin of $001$ topological sector, the result is $h_a\approx0.333332$.
 		And the inset is the density difference
 		$\Delta\langle\hat{n}_i \rangle = \langle\hat{n}_i \rangle_0-\langle\hat{n}_i \rangle_a$ between different topological sector.
 		Here, we choose the Haldane pseudopotential $v_1=1.0$ and circumference of cylinder $L_y=20\ell$.
 	}\label{Afig:pseudo_topological_spin}
 \end{figure}
The topological shift will change as Landau level index, but the difference $2n$ is exactly the
difference of Landau level degeneray on sphere. Please note the degeneracy of $n$th Landau level on sphere is $N_o=2 s_0+1+2n(n=0,1,2,\cdots)$\cite{Haldane1983,Greiter2011}, where $s_0$ is the number of monopole on the center of the sphere and $N_\phi=2s_0$ is the number of magnetic Dirac flux quanta through the surface of the sphere. And the guiding center spin does not change as Landau level index, it can be used to characterize the topological order of FQH state in higher Landau level.

\begin{figure}[t]
	\includegraphics[width=0.46\textwidth]{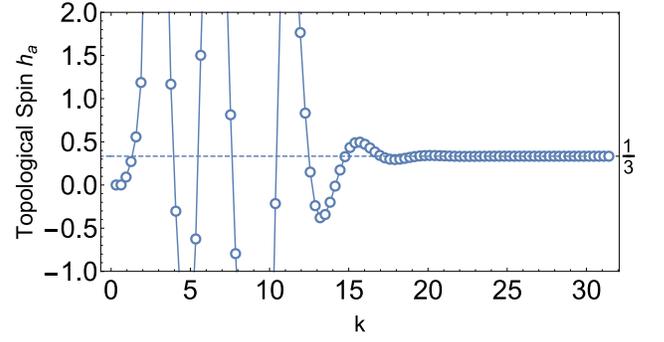}
	\caption{\textit{The ``hardwall edge" of different topological sectors of $\nu=2/3$ state: } The topological spin extracted from density profile, the result is $h_a\approx0.333333$.
		Here, we choose the Haldane pseudopotential $v_1=1.0$ and circumference of cylinder $L_y=20\ell$.
	}\label{Afig:topological_spin_2_3}
\end{figure}

\section{The central charge and topological spin in dipole moment}\label{apdx::modular}

\begin{figure*}
	\includegraphics[width=0.87\textwidth]{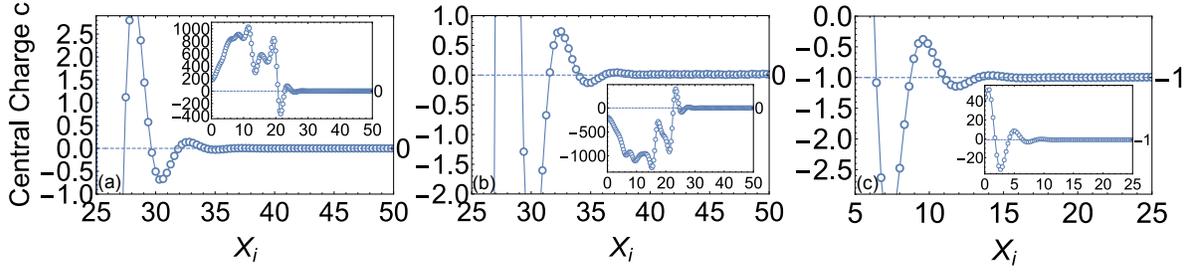}
	\caption{\textit{The central charge $c$ of interface: }
		The central charge of (a)$\nu=1|\nu=1/3$ interface,
		(b)$\nu=0|\nu=2/3$ interface, and (c)$\nu=1|\nu=2/3$ interface.
		The results are $c\approx 0.000031$, $c\approx 0.013280$ and $c\approx-0.996121$ respectively.
	}\label{Afig:Interface_pseudo_cc}
\end{figure*}
In this appendix, we will show more numerical details about the $O(L_y^{-2})$ term in Eq.(\ref{EQ::dipole_MP}).
	
We first consider the dipole moment of ``Hardwall edge" of vacuum sector($h_a=0$), the central charge $c$ can be extracted by
\begin{equation}\label{apdxEQ::cc}
c=24
\left(  \sum_i (\langle\hat{n}_i \rangle-\nu)i+\frac{L_y^2}{8\pi^2}\frac sq\right) +\nu
\end{equation}
using this equation, we have plotted the results of $\nu=1/3,2/3$ states in Fig.\ref{Afig:centralcharge}. The extracted central charge of $\nu=1/3,2/3$ states are $c=0.999931,0.000066$ respectively, in excellent agreement with theoretical predictions.

So far, we have only cut the infinity cylinder into vacuum sector,
now we focus on the quasiparticle sector which includes non-zero topological spin. Motivated by this intuition, we need to cut the infinite cylinder into different topological sectors. For example, the Laughlin $\nu=1/3$ state have $3$ distinct topological sectors which can be constructed by cutting the infinite cylinder $\cdots 010010 \cdots$ into $|010010 \cdots$,$|10010 \cdots$ and $|0010 \cdots$ respectively, where ``$|$'' denotes the ``hardwall edge" and ``$\cdots$" denotes the semi-infinite cylinder. We denote the three setors as $010$, $100$ and $001$. Using the same method as in Sec.\ref{cons_interface} to optimize the MPS, and we have shown the results in Fig.\ref{Afig:pseudo_topological_spin}(a). From the density integral in Fig.\ref{Afig:pseudo_topological_spin}(b), we can clearly see there is a quasiparticle of $100(001)$ sector with charge $Q_a=e/3(-e/3)$. In
Fig.\ref{Afig:pseudo_topological_spin}(c), we have found the dipole moment of $001$ sector is smaller than $010$(vacuum sector). The difference comes from the non-zero topological spin $h_a$,
using Eq.(\ref{EQ::dipole_MP}), the topological spin can be extracted from the density difference between vacuum($\langle\hat{n}_i \rangle_0$) and quasiparticle($\langle\hat{n}_i \rangle_a$) sector
 by
\begin{equation}
h_a = \sum_i (\langle\hat{n}_i \rangle_0-\langle\hat{n}_i \rangle_a)i
\end{equation}
Using this relation, we have shown the results of $\nu=1/3,2/3$ in Fig.\ref{Afig:pseudo_topological_spin}(d) and Fig.\ref{Afig:topological_spin_2_3} respectively, the extrated topological spin are $h_a\approx0.333332$ and $h_a\approx0.333333$, both results are consistent with theorectical predictions.

Finally, we consider the interface version of Eq.(\ref{apdxEQ::cc}). Using Eq.(\ref{EQ::dipole_MP}) and Eq.(\ref{interface:dipole}), we have
\begin{equation}\label{apdxEQ::interface_cc}
c=24
\left(  \sum_i (\langle\hat{n}_i \rangle-\nu_i)i+\frac{L_y^2}{8\pi^2}
\left(\frac{s^R}{q^R}-\frac{s^L}{q^L} \right)   \right)+\nu^R-\nu^L
\end{equation}
where $c$ is the central charge of the interface CFT. In Fig.\ref{Afig:Interface_pseudo_cc}, we have plotted the central charge $c$ of $\nu=1|\nu=1/3$, $\nu=0|\nu=2/3$ and $\nu=1|\nu=2/3$ interface, we have used the same data as in Fig.\ref{fig:IQH_1_3_Interface_pseudo},
Fig.\ref{fig:Vacuum_2_3_Interface_pseudo} and Fig.\ref{fig:IQH1_2_3_Interface_pseudo}.
\textit{(a) $\nu=1|\nu=1/3$ interface: } Following the discussion in Sec.\ref{1|1/3_interfaec}, this interface includes two $1/3$ charged modes and two counter-propagating neutral modes, the total central charge is $0$. In Fig.\ref{Afig:Interface_pseudo_cc}(a), the numerical result is $c\approx0.000031$, consistent with the theoratical value $0$.
\textit{(b) $\nu=0|\nu=2/3$ interface: } The $\nu=0|\nu=2/3$ interface is same as $\nu=1|\nu=1/3$ interface but all chiral edge modes have opposite direction, so the theoretical value of central charge is also $0$. In Fig.\ref{Afig:Interface_pseudo_cc}(b), the extracted central charge is $c\approx 0.013280$.
\textit{(c) $\nu=1|\nu=2/3$ interface: } First, the edge theory of $\nu=2/3$ state is an integer mode and a counter-propagating $1/3$ chiral charged mode. When we put the $\nu=2/3$ state together with $\nu=1$ state, the integer charged mode of $\nu=2/3$ state has been gapped, only remain a single $1/3$ chiral charged mode with opposite chirality(or the edge mode of the hole-type Laughlin $\nu=1/3$ state). So, the central charge of $\nu=1|\nu=2/3$ interface is $c=-1$. In Fig.\ref{Afig:Interface_pseudo_cc}(c), we have shown the central charge and the result is $c\approx-0.996121$.
Both the three numerical results in Fig.\ref{Afig:Interface_pseudo_cc} are consistent with theoretical predictions. Here, we conclude that Eq.(\ref{apdxEQ::interface_cc}) can be used to calculate the central charge of interface.

\end{appendices}

\newpage
\bibliography{ref}

\begin{thebibliography}{81}
\bibitem [{note2()}]{note2}%
  \BibitemOpen
  \bibinfo {note} {Please note the total central charge of $\nu=8/3$ state is
  $c=2$, since the full filled LL of $\nu=8/3$ state include two integer chiral
  edge modes. In maintext, we only consider the $\nu=8/3$ state in the SLL, so
  we can not see the integer chiral edge modes. Conclusively, the central
  charge of $\nu=8/3$ state in the SLL is $c=0$. By the way, the filling factor
  in Eq.(\ref{EQ::MP}) should be $\nu=2/3$.}\BibitemShut {Stop}%
\end{thebibliography}
\newpage
\end{document}